\documentclass[aps,fleqn,showpacs,superscriptaddress,amsmath,amssymb,floatfix]{cas-dc}

\usepackage[numbers]{natbib}
\usepackage{microtype}
\usepackage{graphicx}
\usepackage{dcolumn}
\usepackage{bm}
\graphicspath{ {./images/} }
\usepackage{color}
\usepackage{hyperref}
\usepackage{footnote}
\usepackage{wrapfig}
\usepackage{subcaption}
\usepackage[T1]{fontenc}

\usepackage{amsmath}
\usepackage{amssymb}
\usepackage{amsthm}
\usepackage{amsfonts}
\usepackage{braket}
\usepackage{tabularx} 

\begin{document}
\let\WriteBookmarks\relax
\def\floatpagepagefraction{1}
\def\textpagefraction{.001}
\shorttitle{Ab-initio study of the electronic structure and magnetic properties of Ce$_2$Fe$_{17}$}
\shortauthors{Vishina {\it et al}}

\title [mode = title]{Ab-initio study of the electronic structure and magnetic properties of Ce$_2$Fe$_{17}$}                      

\author[1]{Alena Vishina}[orcid=0000-0002-4583-2877]
\cormark[1]
\ead{alena.vishina@physics.uu.se}

\author[1,2]{Olle Eriksson}
\address[2]{School of Science and Technology, {\"O}rebro University, SE-701 82 {\"O}rebro, Sweden}

\address[1]{Department of Physics and Astronomy, Uppsala University, Box 516, SE-75120, Uppsala, Sweden}

\author[3,1]{Olga Yu. Vekilova}
\address[3]{Department of Materials and Environmental Chemistry, Stockholm University, 10691 Stockholm, Sweden}

\author[1]{Anders Bergman}

\author[1]{Heike C. Herper}

\cortext[cor1]{Corresponding author}

\begin{abstract}
The Ce$_2$Fe$_{17}$ intermetallic compound has been studied intensely for several decades; its low-temperature state is reported experimentally either as ferromagnetic or antiferromagnetic by different authors, with a measured ordering temperature ranging within a hundred Kelvin. The existing theoretical investigations overestimate the experimental total magnetic moment of Ce$_2$Fe$_{17}$ by 20-40 \% and predict a ferromagnetic ground state. By means of first-principle electronic structure calculations, we show that the total magnetic moment of Ce$_2$Fe$_{17}$ can be reproduced within the Local Density Approximation  while functionals based on the Generalized Gradient Approximation  fail. Atomistic spin dynamics  simulations are shown to capture the change in the magnetic state of Ce$_2$Fe$_{17}$ with temperature, and closely replicate the reported helical structure that appears in some of the experimental investigations.
\end{abstract}

\begin{graphicalabstract}
\includegraphics[scale=0.33]{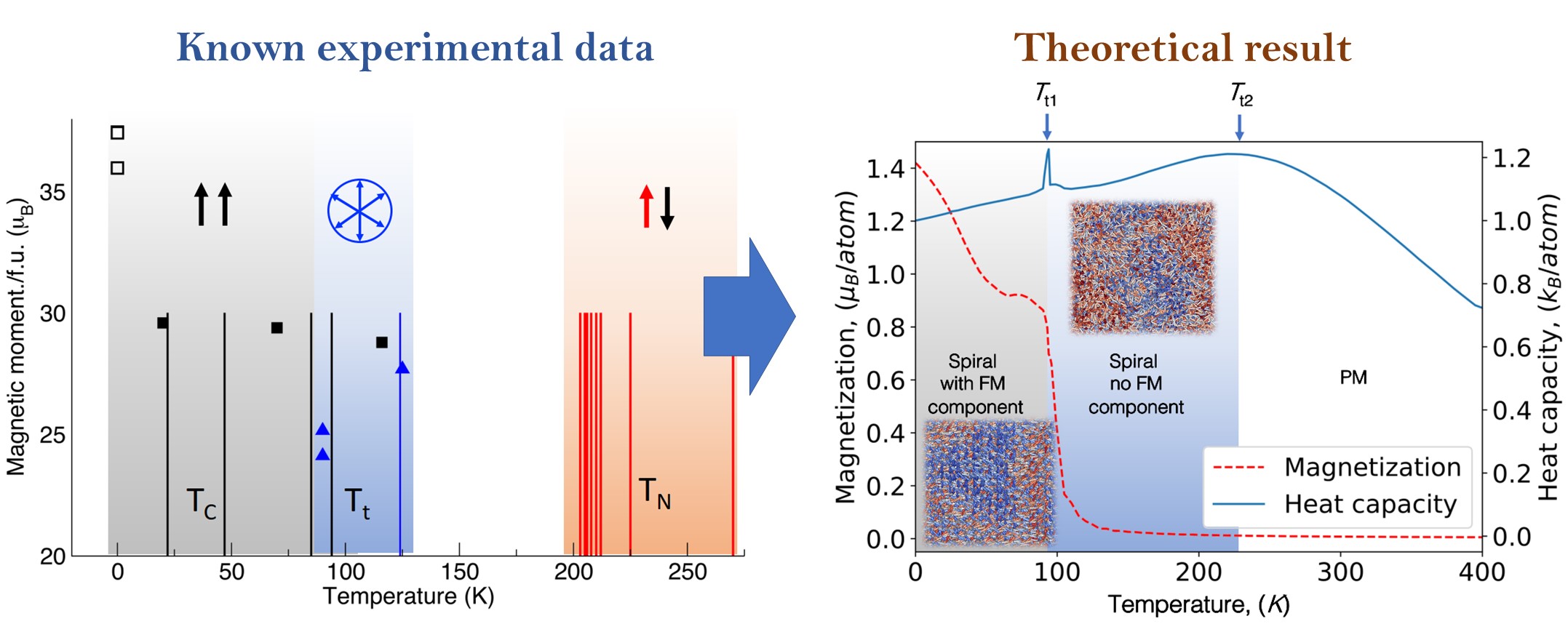}
\end{graphicalabstract}



\maketitle

\section{Introduction}

The Ce$_2$Fe$_{17}$ compound with its high magnetization and low cost can be considered as an attractive material for rare-earth (RE) lean permanent magnets. The drawback of this compound is the low Curie temperature and its basal plane magnetocrystalline anisotropy. A large amount of work has been done to remedy these shortcomings, e.g.  increasing  the magnetic transition temperature  by  nitrogenation or hydrogenation \cite{Isnard_1996,Miraglia}  
as well as replacing  iron partially with Ga \cite{Long}, Al \cite{Mishra}, or Si~\cite{Middleton}. 
It encourages the attempts to tune the properties of this material with the ambition to enhance its use in practical magnetic applications.
A recent theoretical work by Pandey and Parker \cite{PhysRevApplied.13.034039} predicts that Zr and Co  doping of Ce$_2$Fe$_{17}$   improves the magnetic characteristics considerably, making these systems good candidates for permanent-magnetic materials.

Moreover, Ce$_2$Fe$_{17-x}$Z$_{x}$ alloys (with Z = H, C, N, Mn, Ga) show a number of interesting properties such as complex temperature dependence of the magnetocaloric effect in Ce$_2$Fe$_{17}$Mn$_{x}$ \cite{KUCHIN20116763,IWASIECZKO2013130}, giant magnetoresistance \cite{Janssen} in Ce$_2$Fe$_{17}$, spontaneous magnetostriction \cite{ANDREEV200043}, and significant deviation in the lattice volume of Ce$_2$Fe$_{17}$ from a line that shows the monotonic decrease in RE$_2$Fe$_{17}$ unit cell volume from one rare-earth element to the next \cite{Coey}. 

Hence, Ce$_2$Fe$_{17}$ has a large potential for magnetic applications. Theoretical studies of the influence of doping  could  prove even more effective and feasible than the experimental tryouts, provided the properties of the material can be correctly described.
Ce$_2$Fe$_{17}$, along with the different structures of Ce$_2$Fe$_{17-x}$Z$_{x}$ type, have been studied experimentally since the 1970'ies. However, the results of these studies regarding the magnetic state of Ce$_2$Fe$_{17}$ vary a lot and sometimes contradict each other.
We would like to stress, that to our knowledge no conclusive theoretical description of the magnetic properties of Ce$_2$Fe$_{17}$ has been provided that could shed any light on the seemingly controversial experimental results.

This work aims to carefully investigate the best way of describing the electronic and magnetic structure of Ce$_2$Fe$_{17}$. Conclusions made from the present work will hopefully aid in further investigations of the electronic and magnetic properties of various modifications of Ce$_2$Fe$_{17-x}$Z$_{x}$. We start with a brief overview (Sec.~\ref{sec:exp}) of the existing theoretical and experimental data. 
Then we analyze the electronic and magnetic ground state of Ce$_2$Fe$_{17}$ in Sec.~\ref{sec:ground}. By coupling to a low energy spin-Hamiltonian, we also address the magnetic structures at higher temperatures (Sec.~\ref{sec:asd}). Overall, our results show that experimental findings are reproduced and explained both at low and higher temperatures.

\section{Overview: previous theory and experiment}\label{sec:exp}

In this section, we present a short outline of the existing experimental and theoretical data available for Ce$_2$Fe$_{17}$. A more detailed overview can be found in {\it Appendix~\ref{sec:details}}.

\begin{figure}[h!]
\centering
 \includegraphics[scale=0.3]{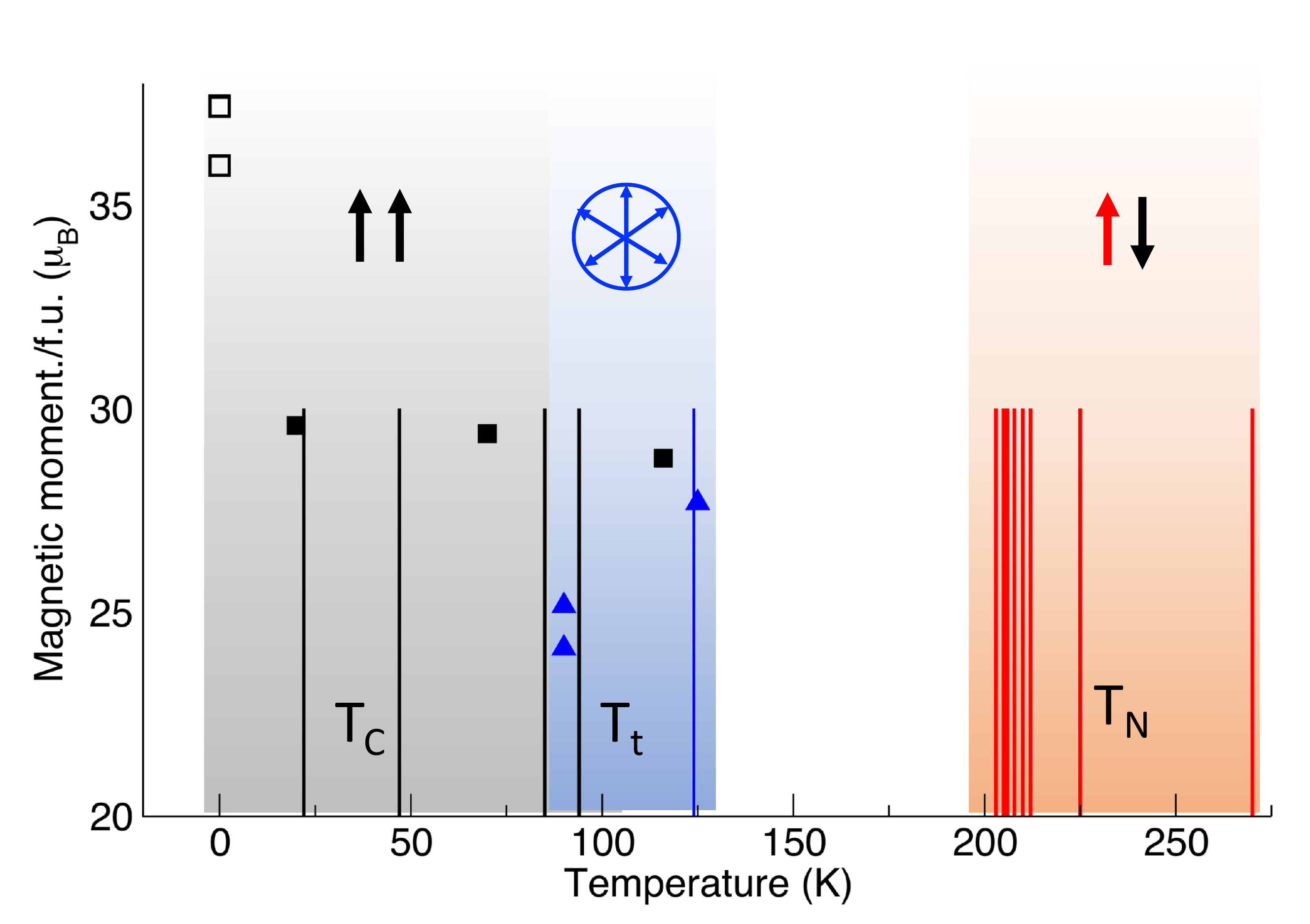}
 \caption{(Color online) Sketch of  magnetic transition temperatures obtained for  Ce$_2$Fe$_{17}$ together with magnetic moments per formula unit (f.u.) if provided (filled squares) in the publications.
 Open squares denote calculated magnetic moments from DFT (T = 0 K). Since no transition temperatures were given in the theoretical work the data were placed at $T=0K$.  Colors and arrows mark the respective magnetic configuration: FM, helical/fan, AFM from the left to the right.
Note: Measured magnetic moments were provided either as moment per Fe atom or per f.u. For the graph moment per f.u. was chosen i.e. if only the moment per Fe atom was available the total magnetic moment was approximated by multiplying by 17 and subtracting 2*0.6$\mu_{\rm B}$ to account for the Ce moment. All data including references can be found in the appendix (Table~\ref{table:1}).}
 \label{Exp}
\end{figure}

The main message we would like to deliver here is that the low-temperature magnetic state and transition temperatures Ce$_2$Fe$_{17}$ reported experimentally vary considerably, depending on sample preparation and the measurement techniques used to determine the magnetic properties. Nonetheless, some general trends can be extracted. We compiled most of them in the diagram in Fig.~\ref{Exp}. 

At low temperatures, a number of experimental papers report  a  ferromagnetic (FM) state \cite{Buschow,kpfu,KUCHIN20007,TEPLYKH2004E99} with  transition temperatures varying between 20 and 160~K and an average moment per Fe atom of 1.8~$\mu_{\rm B}$~\cite{Buschow}. 
A transition to the paramagnetic (PM) state takes place between 206 and 270 ~K \cite{KUCHIN20007,Janssen,Buschow}. The region between the two phases is even more controversially discussed (reporting antiferromagnetic (AFM), non-collinear (NC), and direct transition to PM state, for the details see {\it Appendix~\ref{sec:details}}). 



Some experiments, however, show the absence of the FM phase. Givord and Lemaire \cite{Givord} reported a fan magnetic order of Ce$_2$Fe$_{17}$ below 90 K and a helical magnetic order between T$_{\text{C}}$=90 K and T$_{\text{N}}$=225 K. The fan magnetic structure is parallel to the basal plane with the fan angle of 40$^{\circ}$ at 4.2 K. The helical spin configuration at 140 K is characterized by the rotational axis parallel to {\it c}-axis and magnetization in the basal plane rotating by an angle of 27$^{\circ}$ from plane to plane, and a wave vector $q=0.037$ $\mathrm{\mathring{A}}$; the mean iron moment is 1.55 $\mu_B$.


Janssen {\it et al} \cite{Janssen} observe two transition temperatures, 125 K and 215 K. These are considered to be first the critical temperature at which the magnetic structure changes from one AFM order to another and secondly the Ne{\'e}l  temperature. The authors also state that the first AFM order is not a fan structure  because this phase does not have any FM component. Induced FM order is reported in the higher fields in which the average Fe moment is 1.70 $\mu_B$ at 5 K. 

A helical spin structure with a wave vector $q=0.0372$ $\mathrm{\mathring{A}}$ parallel to the $c$-axis is reported by Fukuda {\it et al} \cite{Fukuda1999} in the temperature region between T$_{\bf C}$=125 K and T$_{\bf N}$=210 K. A modified helix is observed below T$_{\bf C}$ with $q=0.0435$ $\mathrm{\mathring{A}}$.
Depending on the type of the NC structure obtained experimentally, the magnetic moments per Fe atom vary significantly, see Fig.~\ref{Exp} and {\it Appendix~\ref{sec:details}} for details.


A number of authors show a significant decrease in T$_{\text{C}}$ (up to the total suppression of the FM state) and saturation  magnetization  when external pressure is applied to samples \cite{pressure1,pressure2}.

Another aspect, extremely relevant for theoretical description of the system, is the nature of the 4$f$ state of Ce which determines the level of theory needed to describe the electronic structure of Ce$_2$Fe$_{17}$ correctly. 
From the lattice constants Buschow and Wieringen \cite{Buschow} suggested Ce to be tetravalent whereas a mixed valence state of Ce has been reported in Ce$_2$Fe$_{17}$ with the 4{\it f} states being delocalized \cite{Miraglia,Valeanu}.
 A number of authors find hybridization between the Fe 3{\it d}- and Ce 4{\it f}-electron states to be extremely important for the magnetic state of Ce$_2$Fe$_{17}$ \cite{pressure1,pressure2} and suggest that there is no pure localized 4{\it f}$^1$ configuration of Ce \cite{Coey,PhysRevLett.60.2523} in this material.  

Furthermore,  the 4{\it f} electron character might change  as the effect of doping~\cite{FUJII199510,Fujii,ISNARD1992157,Miraglia,Valeanu,ISNARD1997198} (for the details see {\it Appendix~\ref{sec:details}}). It is shown that a considerable increase in volume, magnetization, and T$_{\bf C}$ is observed along with the decrease in Ce valence i.e. increase in localization \cite{Miraglia,Isnard_1996,ISNARD1997198,Hybrid}.


The existing theoretical research into Ce$_2$Fe$_{17}$ basically fails to reproduce the experimental magnetic state of the compound. The reported total magnetic moment considerably overestimates the experimental values (of around   28 $\mu_B$/f.u. at low temperatures).
Using the Tight-Binding Linear Muffin-Tin Orbitals Atomic-Sphere Approximation (TB-LMTO-ASA)   \cite {Andersen} without spin-orbit coupling and in Local Density Approximation (LDA) Lukoyanov {\it et al.} obtained a moment of 35.98 $\mu_B$/f.u.
An even higher moment (38.04 $\mu_B$/f.u.) has been reported in the recent Full-potential Density Functional Theory (DFT) study within the Generalized Gradient Approximation (GGA) and the Hubbard U correction for the Ce 4{\it f} states by Pandey and Parker \cite{PhysRevApplied.13.034039}. The same overestimation was obtained in Ref.~\cite{PhysRevMaterials.3.084407} for a plain DFT/GGA calculations using the VASP code \cite{vasp1,vasp2,vasp3,vasp4}  and the Projector Augmented Wave (PAW) \cite{PAW} method. 
All above mentioned studies assumed a collinear FM ground state. For details see {\it Appendix~\ref{sec:details}}.

The main conclusion from this section is that there exists not only some controversy in experimental results obtained for Ce$_2$Fe$_{17}$, but also a large discrepancy between the theoretical and experimental data, see Table \ref{table:1}. Moreover, it is difficult to identify any theoretical work explaining the reason for these contradictions. The motivation of the present study is to provide an explanation to all these confusing results, and to provide a microscopical understanding of the magnetic properties of Ce$_2$Fe$_{17}$.  

\section{Computational methods}\label{sec:method}

The relaxed crystal structure of Ce$_2$Fe$_{17}$ was determined using Vienna Ab Initio Simulation Package (VASP) \cite{vasp1,vasp2,vasp3,vasp4} within the Projector Augmented Wave (PAW) method \cite{PAW}. For Ce the potential with 5{\it s}$^2$5{\it p}$^6$4{\it f}$^1$5{\it d}$^1$6{\it s}$^2$ valence configuration was chosen when the {\it f} electron was treated as a valence state; a 5{\it s}$^2$5{\it p}$^6$5{\it d}$^1$6{\it s}$^2$ configuration was used to treat 4{\it f}$^1$ as a core state. LDA was used in Ceperley and Alder form \cite{Ceperley} in the parametrization of Perdew and Zunger \cite{PerdewZunger}. Also, the 
GGA in Perdew, Burke, and Ernzerhof (PBE) form \cite{PBE} was employed, as well as the LDA+{\it U} correction scheme \cite{Anisimov,Solovyev}. The plane-wave energy cutoff was set to 400 eV.

DFT + {\it U} calculations were performed in VASP within the approach introduced by Dudarev {\it et al} \cite{Dudarev} which depends only on the effective $U_{eff}=U-J$ with Hubbard parameter $U$ and Hund's coupling parameter $J$.
For the non-collinear VASP calculation in the conventional (57 atoms/u.c.) and double (117 atoms/u.c.) unit cells, the integration of the Brillouin zone was performed using 3x3x3 and 3x3x1 Monkhorst-Pack \cite{Monkhorst} k-point sampling, respectively.
Fixed spin moment (FSM) calculations were conducted with VASP by performing the volume-relaxation (c/a ratio and  shape unchanged) for each value of the fixed total magnetic moment. These calculations were performed in GGA in the PBE form.

The unit cell magnetization was also calculated with the full-potential linear muffin-tin orbital method (FP-LMTO), including spin-orbit interaction as implemented in the RSPt code \cite{rspt1, rspt2}, with the PBE functional \cite{PBE} for exchange and correlation. The results were obtained with the tetrahedron method with Bl\"ochl correction for the Brillouin zone integration \cite{Blochl}; the converged k-point Monkhorst-Pack meshes \cite{MP}  $24\times24\times24$ were used for the calculations. RSPt was also used to calculate the interatomic exchange interactions as well as the hybridization function $\Delta (E)$ with an additional iteration on top of the converged calculations.

 The Curie temperature was determined using Metropolis Monte Carlo (MC), and Atomistic Spin Dynamics (ASD) simulations implemented within the Uppsala  Atomistic  Spin  Dynamics (UppASD) software \cite{ASD}. The simulations were performed on a $32\times 32\times 32$ supercell with periodic boundary conditions using the  exchange parameters calculated with the RSPt code within the first nine coordination shells and with the PBE functional \cite{PBE} for exchange and correlation. From the MC simulations, the critical temperatures were determined by the location of divergences in the magnetic heat capacity $C_m$, calculated as $C_m = \frac{d<U>}{dT}$ where $<U>$ is the average total energy per spin as obtained from the MC simulations.  
 
The presence of spin-spirals in the simulated magnetic structures is identified by means of the static structure factor  $S(\mathbf{q})$, i.e. the Fourier transform of the equal-time spatial correlation function  $C(\mathbf{r})$\cite{ASD}. Wave-vectors $\mathbf{q}$ that exhibit a large spectral weight in $S(\mathbf{q})$ indicates the presence of spin-spirals with the same wave-vector $\mathbf{q}$ in the sample. Note that that using this convention, $S(\mathbf{q}=\mathbf{0})$ corresponds to the average magnetization $m$ and is thus non-zero if the magnetic structure contains a ferromagnetic component.
 
 The Sumo package \cite{sumo} was used for the density of states (DOS) plots.

\section{Results}\label{sec:main}


\subsection{Ground state with DFT}\label{sec:ground}

\begin{figure}[h!]
 \includegraphics[scale=0.32]{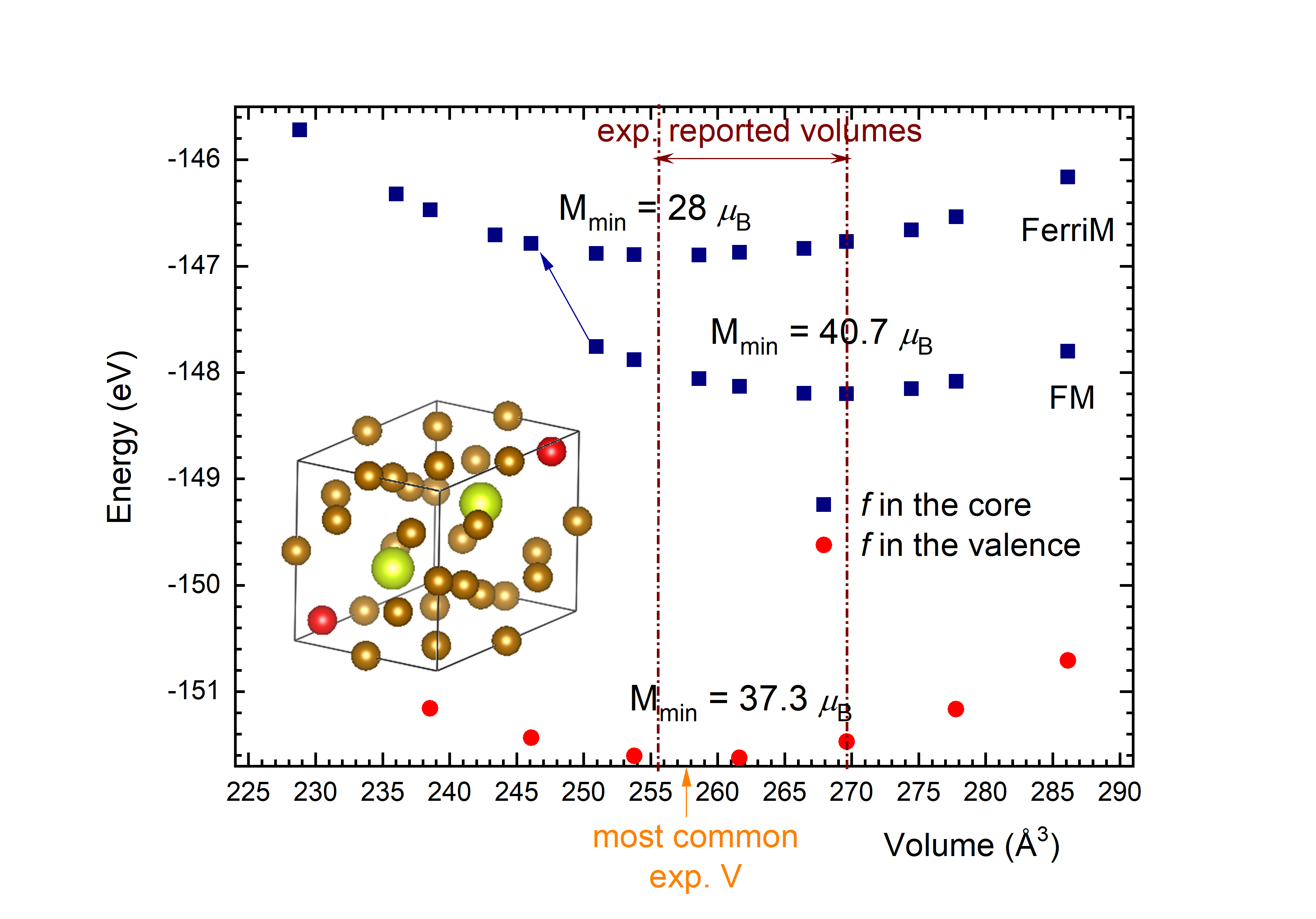}
 \caption{(Color online) Calculated total energy versus unit cell volume, for Ce$_2$Fe$_{17}$, with 5{\it s}$^2$5{\it p}$^6$4{\it f}$^1$5{\it d}$^1$6{\it s}$^2$ valence configuration (red circles) and 5{\it s}$^2$5{\it p}$^6$5{\it d}$^1$6{\it s}$^2$ valence configuration (blue squares). The latter curve has a FerriM and FM branches; each curve has the magnetic moment of the relaxed state written above it. The range of experimental volumes found in ICSD database \cite{ICSD} along with the most common experimental volume are also given in the graph. The GGA functional was used for all calculated data-points. The figure also shows the primitive unit cell of $\beta$-Ce$_2$Fe$_{17}$ (inset). Iron atoms are shown with  brown spheres, two Fe atom in 6{\it c} positions are highlighted with red, Ce atoms are yellow.}
 \label{Vrelax}
\end{figure}

$\beta$-Ce$_2$Fe$_{17}$ crystallises in R-3{\it m} crystallographic space group with two cerium atoms occupying the 6{\it c} site; two iron atoms are situated in 6{\it c} positions, 3 Fe atoms are occupying 9{\it d} sites, and six Fe atoms are in each of 18{\it f} and 18{\it h} sites (see inset of  Figure \ref{Vrelax}).


The current investigation started with a crystal structure relaxation, performed using VASP. 
The nature of 4{\it f} states in Ce (localized or itinerant, or something in between) varies depending largely on the compound, cf Sec.~\ref{sec:exp} . 
To investigate the effect of the choice of the valence state we performed unit cell relaxation with two different valence configurations: 5{\it s}$^2$5{\it p}$^6$4{\it f}$^1$5{\it d}$^1$6{\it s}$^2$ was chosen when the 4{\it f} electron was treated as a valence state; 5{\it s}$^2$5{\it p}$^6$5{\it d}$^1$6{\it s}$^2$ configuration was used to treat 4{\it f}$^1$ as a core state. The former corresponds to an itinerant 4{\it f} electron state of Ce, while the latter is appropriate for the localized 4{\it f}-state. The total energy, calculated in GGA, is shown in Figure \ref{Vrelax} as a function of the volume; the relaxed unit cell parameters and corresponding magnetic moments are given in Table \ref{table:2}.

The experimental unit cell volumes of Ce$_2$Fe$_{17}$ are given in the Inorganic Crystal Structure Database (ICSD)  \cite{ICSD} and are found to vary between 255.7 and 269.6 $\mathrm{\mathring{A}}^3$, although the most commonly occurring volumes fall between 258 and 259 $\mathrm{\mathring{A}}^3$. Treating the 4{\it f} electron as a core state leads to a minimum of the unit cell volume of 267.82 $\mathrm{\mathring{A}}^3$, which overestimates the typical experimental value. In the relaxed state, Fe atoms are oriented ferromagnetically (here we define FM as the parallel orientation of magnetic moments of all Fe atoms, even though the moments of the Ce atoms are oriented antiferromagnetically to the Fe spins) with a total magnetic moment of 40.65 $\mu_B$/f.u. (Table \ref{table:2}). This is a moment that excludes the 4f moment of Ce. 
In the case of a free atom, for a  trivalent Ce ion (Ce$^{3+}$) the 4{\it f} angular momentum is $J=L-S = 5/2$, and the Land\'e factor is given by $g_J =6/7$, such that the total moment contributed by the Ce 4{\it f} electron amounts to 2.14~$\mu_{\rm B}$. This moment is found to be directed parallel to the spin moment of Fe (anti-parallel to the Ce {\it spd}  moment). For this reason a total moment of 44.93 $\mu_B$/f.u. can be expected for the 4{\it f} as core calculation, which is much too large compared with the experimental results. Hybridization function calculations described further show, however, that this treatment of Ce 4{\it f} moment is not entirely correct as Ce ion is in an intermediate valence state. The calculated atom projected magnetic moments for this calculation are: 2.64 $\mu_B$ for the Fe 6{\it c} site, 2.20-2.22 $\mu_B$ for the Fe 9{\it d} site, 2.46-2.49 $\mu_B$ for the Fe 18{\it f} site, 2.41-2.44 $\mu_B$ for the Fe 18{\it h} site, and -0.31 $\mu_B$ for the Ce 6{\it c} site. We note that the moments of the Fe 9{\it d} site and the Fe 18{\it f} site have range of values, since the atoms have slightly different environment. 
The highest magnetic moment belongs to the 6{\it c} site of Fe which are highlighted with red color in Figure \ref{Vrelax}.

Another interesting feature observed here is the magnetic phase transition which happens as the unit cell of Ce$_2$Fe$_{17}$ is decreased in volume. This is shown with a blue arrow in Figure \ref{Vrelax}. The main difference of the new magnetic state is the magnetic moment of the Ce 6{\it c} site; we define this state as a ferrimagnetic (FerriM) state. The relaxed total moment of the system when it is forced to stay in the FerriM state is 28.01 $\mu_B$/f.u. (-0.35 $\mu_B$ for the Fe 6{\it c} site, 1.28 $\mu_B$ for the Fe 9{\it d} site, 2.10-2.12 $\mu_B$ for the Fe 18{\it f} site, 2.10-2.12 $\mu_B$ for the Fe 18{\it h} site, and -0.26 $\mu_B$ for the Ce 6{\it c} site). Here the magnetic moment of the Fe (6c) sites is basically quenched. Again adding the localized 4f moment of Ce results in a total moment of 32.29 $\mu_B$/f.u. At larger volumes this state is meta stable, but for compressed volumes it represents the lowest energy state (for the electronic configuration with Ce 4f states are treated as core electrons).

Treating the 4{\it f} electron as an itinerant valence state gives the relaxed unit cell volume of 256.92 $\mathrm{\mathring{A}}^3$, which is much closer to the most common experimental one. The calculated total magnetic moment of 37.33 $\mu_B$ is lower than the moment obtained while treating {\it f} electron as a core, but it is still much higher than the experimental value of 28-31 $\mu_B$/f.u. This FM state is characterized by local moments of 2.58 $\mu_B$ for the Fe 6{\it c} site, 2.04-2.08 $\mu_B$ for the Fe 9{\it d} site, 2.38-2.41 $\mu_B$ for the Fe 18{\it f} site, 2.20-2.21 $\mu_B$ for the Fe 18{\it h} site, and -0.80 $\mu_B$ for the Ce 6{\it c} site, which are in accordance with the other theoretical works that employed the GGA functional \cite{PhysRevApplied.13.034039,PhysRevMaterials.3.084407}. Again, the Fe (6c) atoms provide the largest moment also for this calculation.

As the next step, we performed similar calculations using  LSDA treating the Ce 4f electrons as itinerant valence states. The LSDA functional vastly underestimates the u.c. volume, see Table \ref{table:2}, which is common for LSDA. Magnetic moment evaluated for this compressed volume is much smaller than the experimental one. However, the calculated magnetic moment is quite close to the experimental value if one calculates it within LSDA for the experimental unit cell parameters. As we see in Table \ref{table:2}, for the experimental volume we obtain a total magnetic moment of 28.47 $\mu_B$/f.u. (with atomic resolved moments as: 2.30 $\mu_B$/Fe 6{\it c}, 1.32-1.34 $\mu_B$/Fe 9{\it d}, 1.72-1.73 $\mu_B$/Fe 18{\it f}, 1.76-1.78 $\mu_B$/Fe 18{\it h}, and -0.57 $\mu_B$/Ce 6{\it c}). 


The DFT + {\it U} formalism \cite{Anisimov,Solovyev} is often used to describe the (partial) localization of rare-earth 4{\it f} electrons, although we note that several articles have found faults in its applications to the 4{\it f} shell, when it is occupied by more than one electron \cite{PhysRevB.89.205109,PhysRevB.94.085137}. As we mentioned in   Sec.~\ref{sec:exp},  the mixed valence state of Ce is reported in several experimental works on Ce$_2$Fe$_{17}$X-systems. Hence, the method of treating the 4{\it f}-electrons of Ce (either as itinerant or localized) might have an important effect on the magnetic state of Ce$_2$Fe$_{17}$ and should be investigated. We used several values of $U-J$ for the 4{\it f} electron of Ce; as reported in Table \ref{table:2}, this approach makes little change to the total magnetic moment and crystal structure parameters of Ce$_2$Fe$_{17}$ and does not reproduce the experimental value of the total magnetic moment (for the details, including the density of states calculated with LSDA, GGA, and GGA + {\it U} see  Appendix~\ref{sec:DOS}). 

Since some of the experiments (see Sec.~\ref{sec:exp}) show a non-collinear ground state of Ce$_2$Fe$_{17}$, while one of the investigations \cite{Supercell} suggests a double cell superstructure,  we performed non-collinear GGA VASP calculations for the primitive unit cell (with 57 atoms per u.c.) and a supercell that was doubled in the {\it z}-direction (with 114 atoms per u.c.). However, according to our calculations, the collinear orientation of spins is lower in energy, for any of the volumes investigated. 

\begin{table*}[t]
\caption{Magnetic moment per unit cell of Ce$_2$Fe$_{17}$, calculated for the relaxed crystal structures (Fig. \ref{Vrelax}) and the moment calculated for the most common experimental volume (V = 258.17  $\mathrm{\mathring{A}}^3$ \cite{TEPLYKH2004E99})  along with the relaxed crystal structure parameters and T$_C$. The values are calculated with different computational approaches, exchange correlation functionals ($E_{xc}$), and two different descriptions of the Ce 4{\it f} state (localized core and itinerant valence). Note, that for the 4\textbf{f} as core calculation the contribution from the 4{\it f} Ce moment is not included in the listed total moment.}
\begin{tabular}{c c c c c c c c c c c c} 
 \hline \hline
 \multicolumn{3}{c}{Method} & M &  a & c & V & Ordering temperature\\ 
 Code& $E_{xc}$ & $4f$  & $\mu_B$/f.u. & $\mathrm{\mathring{A}}$ & $\mathrm{\mathring{A}}$ & $\mathrm{\mathring{A}}^3$ & K \\
 \hline
\multicolumn{5}{c}{{\bf Relaxed structures}} & &  & & & & \\
VASP& LSDA &valence & 6.3 & 8.042 & 11.611 & 216.78 & \\
VASP& LSDA & $U$=3 eV & 4.4 & 8.055 & 11.697 & 217.62 & \\
VASP& GGA & core & 40.65  & 8.605 & 12.589 & 267.82 &  \\ 
VASP& GGA & valence & 37.68 & 8.415 & 12.649 & 258.56 &  \\
VASP& GGA&  $U$=3 eV & 37.92 & 8.458 & 12.541 & 258.97 &  \\
VASP& GGA&  $U$=5 eV & 37.86 & 8.449 & 12.535 & 258.28 &  \\
\multicolumn{5}{c}{{\bf Most common experimental volume}} & & & & & & \\
VASP& LSDA &valence & 28.47 & & & & \\
VASP& LSDA & $U$=3 eV & 27.37 & & & & \\
VASP& GGA & core & 39.45 & & & &  \\ 
RSPt& GGA& core  & 37.61 & & & & 575 \\ 
RSPt& GGA & valence  & 36.99 & & & & 625 \\
RSPt& LSDA & core  & 32.76 & & & & non-collinear \\
RSPt& LSDA & valence  & 32.24 & & & & non-collinear \\
 \hline \hline
\end{tabular}
\label{table:2}
\end{table*}

\label{table:2}

To crosscheck our findings  full-potential calculations were performed using the FP-LMTO method as implemented in the RSPt code \cite{rspt1, rspt2}  and   the Ce$_2$Fe$_{17}$ structures relaxed in VASP. The two methods give similar results (see Table \ref{table:2}), with a high total magnetic moment, both when treating the 4{\it f} electrons as valence and as a core state. The results from the  RSPt calculations are consistent with our  findings from VASP, namely, Fe in the 6{\it c} site carrying the largest moment while Ce spin moment is directed anti-parallel to Fe.

In the past, it has been shown that the 4{\it f} hybridization function, being a good indicator of 4{\it f} orbital overlap, can help to understand the degree of localization of Ce  4{\it f} states \cite{Hybrid}. Therefore, we calculated the hybridization function, $\Delta (E)$, for the largest and the smallest experimental unit cell volumes found in ICSD (see Fig. \ref{Hybridization},  Appendix~\ref{sec:hybgga}) treating the Ce 4{\it f} states are itinerant. The absolute value of the peak is situated about 1 eV below E$_F$, and it is seen to decrease from 0.27 to 0.23 with the increase in volume (in LSDA). The values themselves, if compared with results for other Ce-based compounds \cite{Hybrid}, show an intermediate magnitude of the hybridization function in the range of $\gamma$-Ce, that is between that of the materials which are known to have itinerant 4{\it f} electrons (like CeO$_2$ \cite{CeO2Ce2O3})) and the materials with a well established, localized 4{\it f} shell (e.g. CeBi\cite{MONOPNIC}). This result for Ce$_2$Fe$_{17}$  supports  the findings by Coey \cite{Coey} who pointed to the intermediate nature of the 4{\it f} state (see Sec.\ref{sec:exp}).   Buschow's argumentation\cite{Buschow} that the 4{\it f} state must be itinerant, was solely based on the volume and that alone is not a sufficient criterion and in case of $3d$ valence electrons of the ligand (Fe) the systems are slightly more localized even for small volumes than systems with e.g. $p$ electrons\cite{Hybrid}. 
(For comparison the hybridization function was also calculated in GGA, see discussion in Appendix~\ref{sec:hybgga}.)

Fixed spin moment (FSM) calculations are sometimes utilized to investigate the existence of the local and global energy minima, as function of the total magnetic moment \cite{FixM}. We performed FSM calculations for Ce$_2$Fe$_{17}$ by changing the total moment within a region of values containing the experimental magnetic moment of about 29 $\mu_B$/f.u. and a calculated magnetic moment of 38 $\mu_B$/f.u. In these calculations the 4{\it f} electrons were treated as valence (GGA, without a {\it U}-term). The details can be found in Appendix~\ref{sec:fixed}. No pronounced minimum is observed around the experimental value, however, the curve does demonstrate a slight deviation from the monotonic decrease in the region around 30 $\mu_B$/f.u.

Hence, neither GGA nor LSDA can reproduce all the observed ground state properties of Ce$_2$Fe$_{17}$ in a consistent way.
While LSDA considerably underestimates the volume of the unit cell GGA reproduces the experimental crystal structure well. Unfortunately, GGA calculations noticeably overestimate the magnetic moment. At the same time, LSDA gives a total magnetic moment very close to the experimental one, if the experimental unit cell parameters are used. Hence, from a practical point of view, if the magnetic properties of Ce$_2$Fe$_{17}$ are of interest, the LSDA functional is to be preferred, in combination with the treatment of the Ce 4{\it f} states that is somewhere between the itinerant and localized. Such an approach could be based on the dynamical mean field theory for the 4{\it f} states of Ce in combination with LSDA for rest of the valence states. Such a study, however, is outside of the scope of the present investigation. 

\subsection{Curie temperature and magnetic structure from atomistic spin dynamics simulations}\label{sec:asd}

As we can see from the overview given in Sec.~\ref{sec:exp}, a number of magnetic configurations have been reported for the low-temperature state of Ce$_2$Fe$_{17}$, including FM, AFM, and non-collinear spin structures. To analyze the magnetic structure at finite temperatures and estimate the Curie temperature, we performed Monte Carlo and atomistic spin-dynamics simulations of Ce$_2$Fe$_{17}$. 

For the magnetic moments and exchange parameters $J_{ij}$ obtained with GGA (both {\it f in the valence} and {\it f in the core}) Ce$_2$Fe$_{17}$ remains FM (for computational details see Sec.\ref{sec:method}) at higher temperatures. The example of a simulated spin structure is given in Appendix~\ref{sec:spinGGA}, Fig.~\ref{FMsnap}. T$_C$ depends on the volume (and the way 4{\it f} is treated) and equals to 460 K for the lowest reported experimental volume (V = 255.77 $\mathrm{\mathring{A}}^3$ \cite{767}) and 575 K for the most common experimental volume (V = 258.17  $\mathrm{\mathring{A}}^3$ \cite{TEPLYKH2004E99}) (4{\it f} is treated as core). These temperatures are much higher than any of the reported in experiments, see Table~\ref{table:1}. Hence, we can conclude that not only does GGA fail to capture the correct magnetic moment at 0 K, but the ASD calculations based on GGA input overestimate the magnetic moment at higher temperatures and produce the Curie temperature at least four times higher than the T$_{\rm C}$ observed experimentally.

The picture changes, however, when the LDA magnetic moments and exchange parameters are used as an input for the ASD simulations. Non-collinear spin structure is observed at low temperatures for all the u.c. volumes of Ce$_2$Fe$_{17}$. All the simulations described further are based on the LDA magnetic configuration, treating the 4f states as core electrons.   
\begin{figure}[h]

\centering
\includegraphics[scale=0.27]{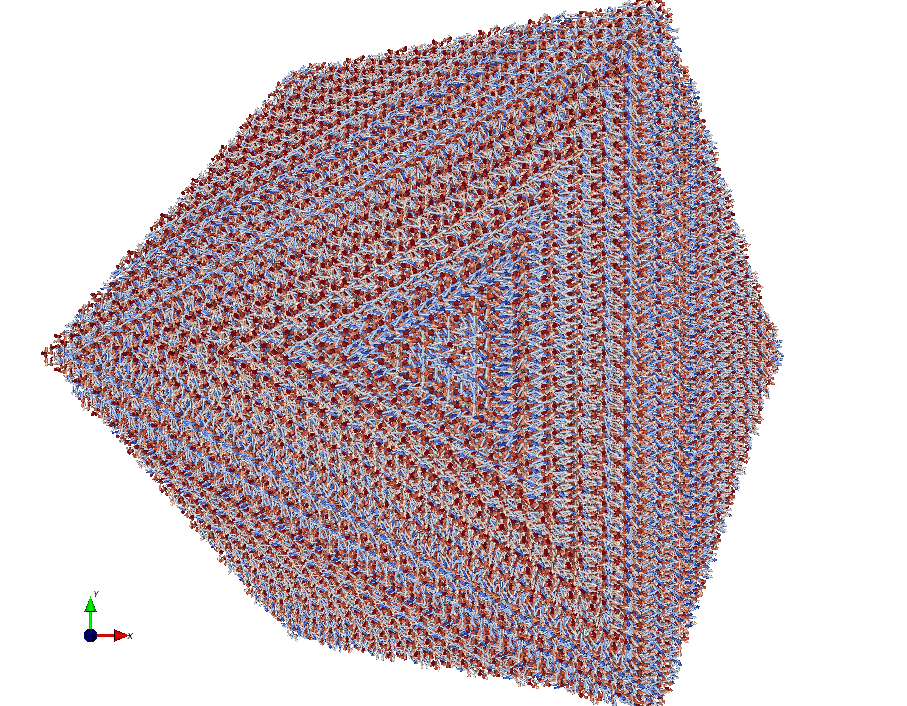} 
 
\caption{(Color online) Magnetic configuration of Ce$_2$Fe$_{17}$ at T = 0 K, obtained from ASD simulations based on exchange parameters calculated in LDA for experimental crystal structure with V = 258.17 $\mathrm{\mathring{A}}^3$ \cite{TEPLYKH2004E99}.}
\label{SpinUpp}
\end{figure}

\begin{figure}[h!]
 \centering
 \includegraphics[scale=0.33]{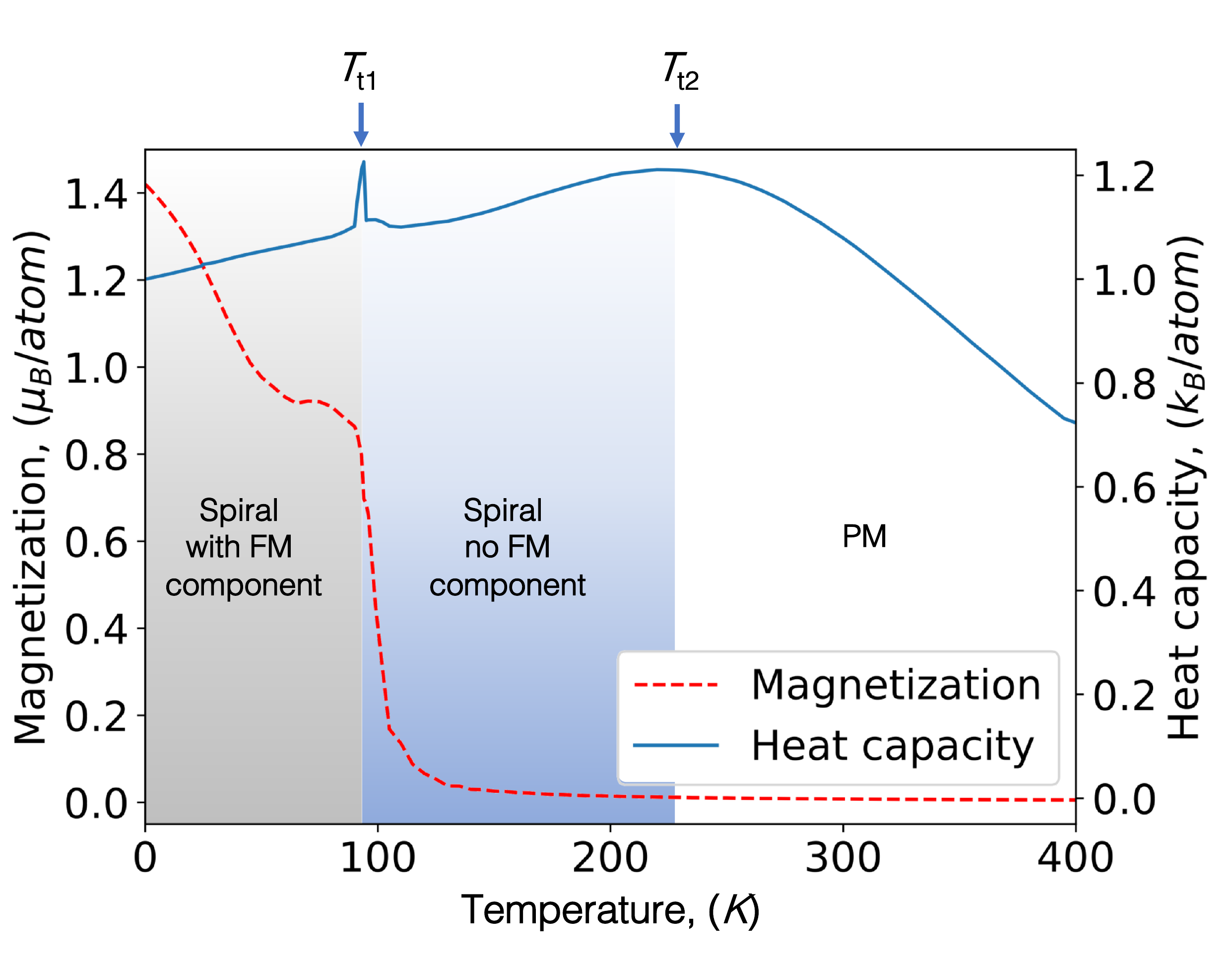}
 \caption{(Color online) The change in specific heat with temperature calculated with UppASD code (unit cell volume was 258.17  $\mathrm{\mathring{A}}^3$). Two peaks in specific heat, show two transition temperatures of Ce$_2$Fe$_{17}$ (see discussion in the text).}
 \label{Cv}
\end{figure}
The low-temperature spin structure, obtained for the most common experimental volume (V = 258.17  $\mathrm{\mathring{A}}^3$ \cite{TEPLYKH2004E99}), is shown in Figure~\ref{SpinUpp}. The net moment for this configuration is around 27 $\mu_B / f.u.$ which agrees well with the experimental data.

Specific heat calculated with UppASD up to the temperature of 400 K catches a phase transition (see the narrow peak in Figure~\ref{Cv} denoted as $T_{\rm t1}$). We observe a spin-spiral state with a significant ferromagnetic component at low temperatures (up to the first transition temperature of T$_{\rm t1}$ = 93 K). This result explains the FM state reported in several experiments at low temperatures and the corresponding magnetic moment, which is noticeably lower than the moment one would expect for seventeen atoms of Fe per unit cell.  The magnetization obtained for the low temperature (Fig.~\ref{Cv}) is close to the experimental (where the FM state is reported). A different spin-spiral state can also be observed above the temperature of 93 K, and this state is stable until the temperature of around 220 K (see the broad peak in Figure~\ref{Cv} denoted as $T_{\rm t2}$). A correlation function analysis in this region $T_{\rm t1}$<T<$T_{\rm t2}$ shows that some $q$-vectors seem to be more populated than others, which is not expected in a paramagnetic phase (data not shown). However, long-range order is no longer obvious and the spin-spiral state in that temperature range is probably due to the short-range order.

\begin{figure*}[hbt!] 
\begin{subfigure}{0.54\textwidth}
\includegraphics[width=\linewidth]{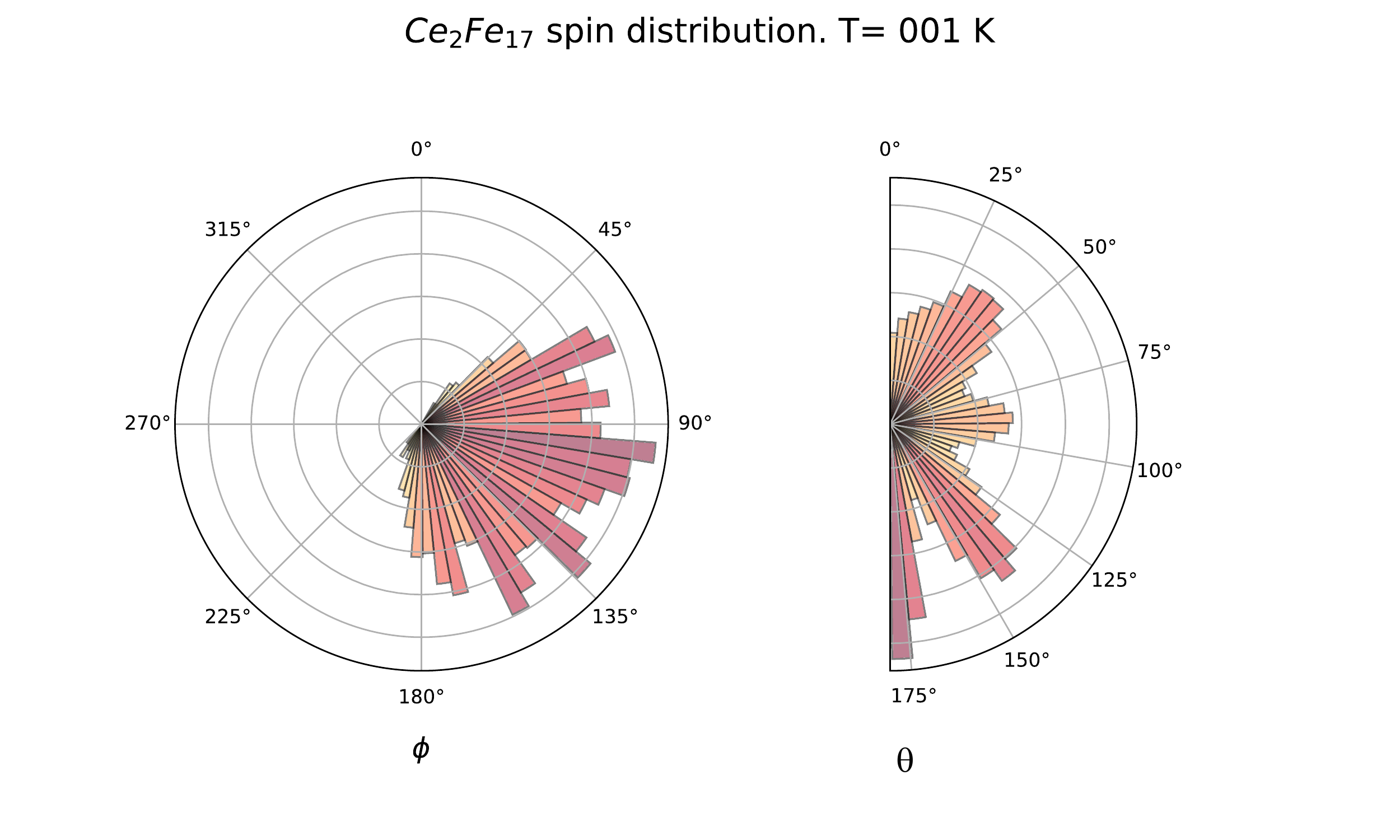}
\end{subfigure}\hspace*{\fill}
\begin{subfigure}{0.54\textwidth}
\includegraphics[width=\linewidth]{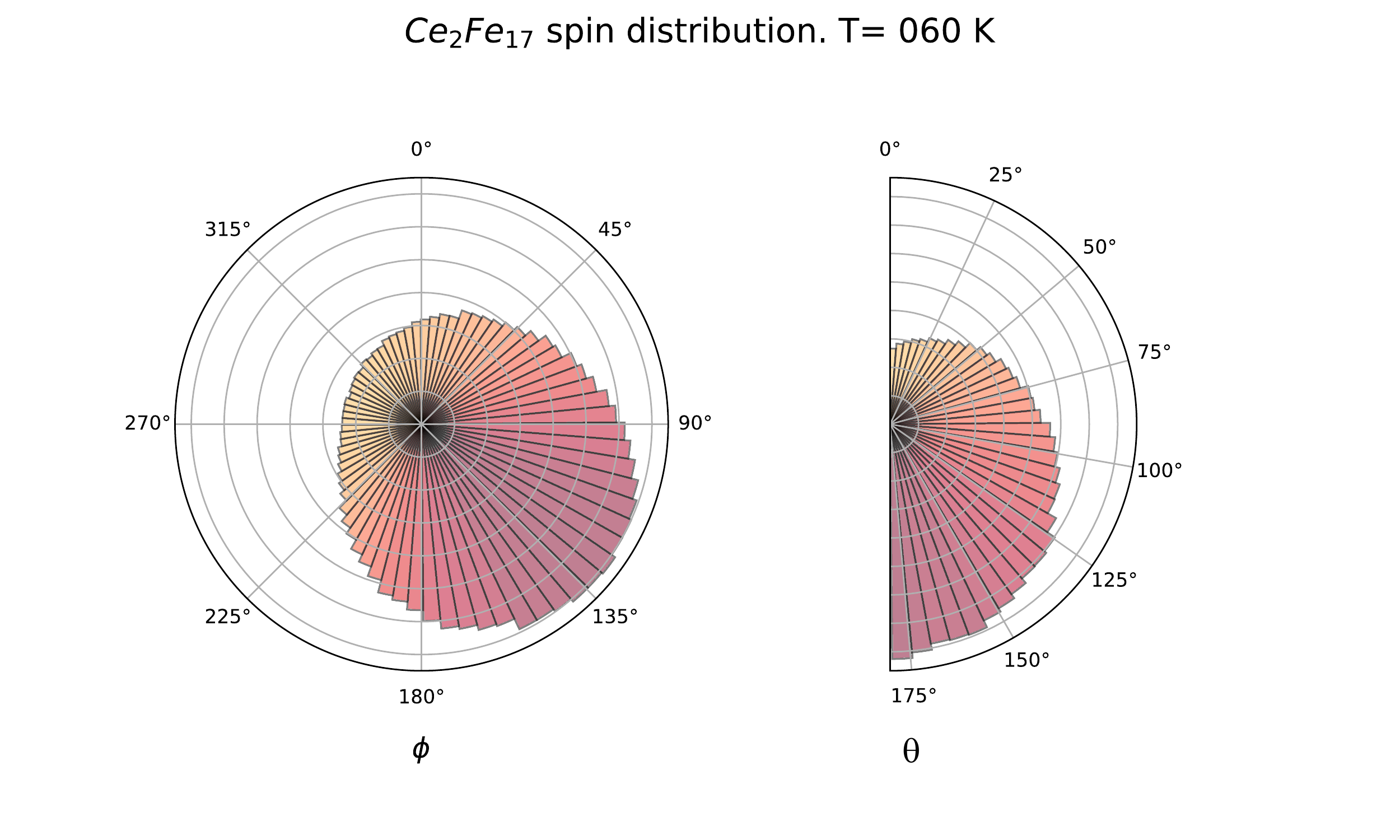}
\end{subfigure}

\medskip
\begin{subfigure}{0.54\textwidth}
\includegraphics[width=\linewidth]{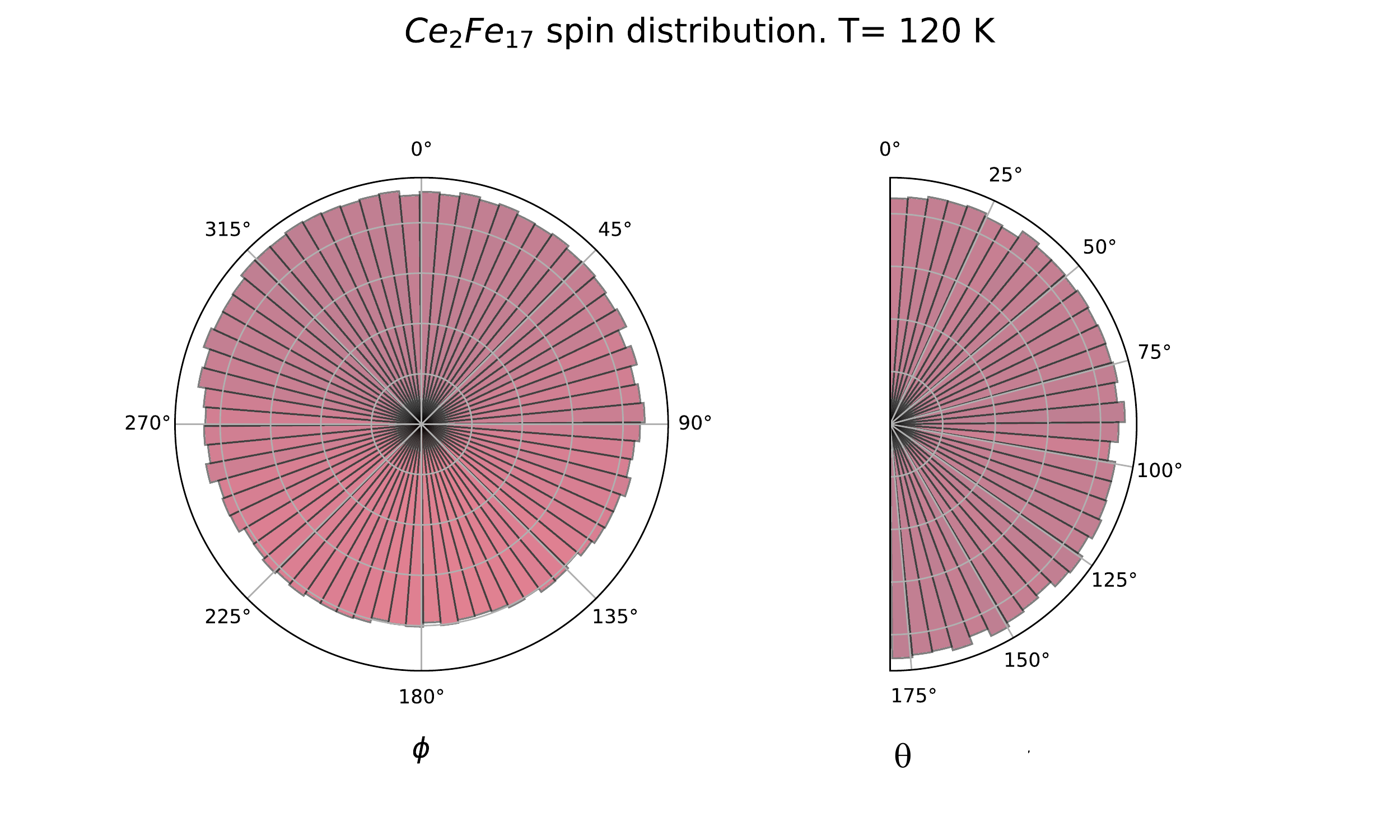}
\end{subfigure}\hspace*{\fill}
\begin{subfigure}{0.54\textwidth}
\includegraphics[width=\linewidth]{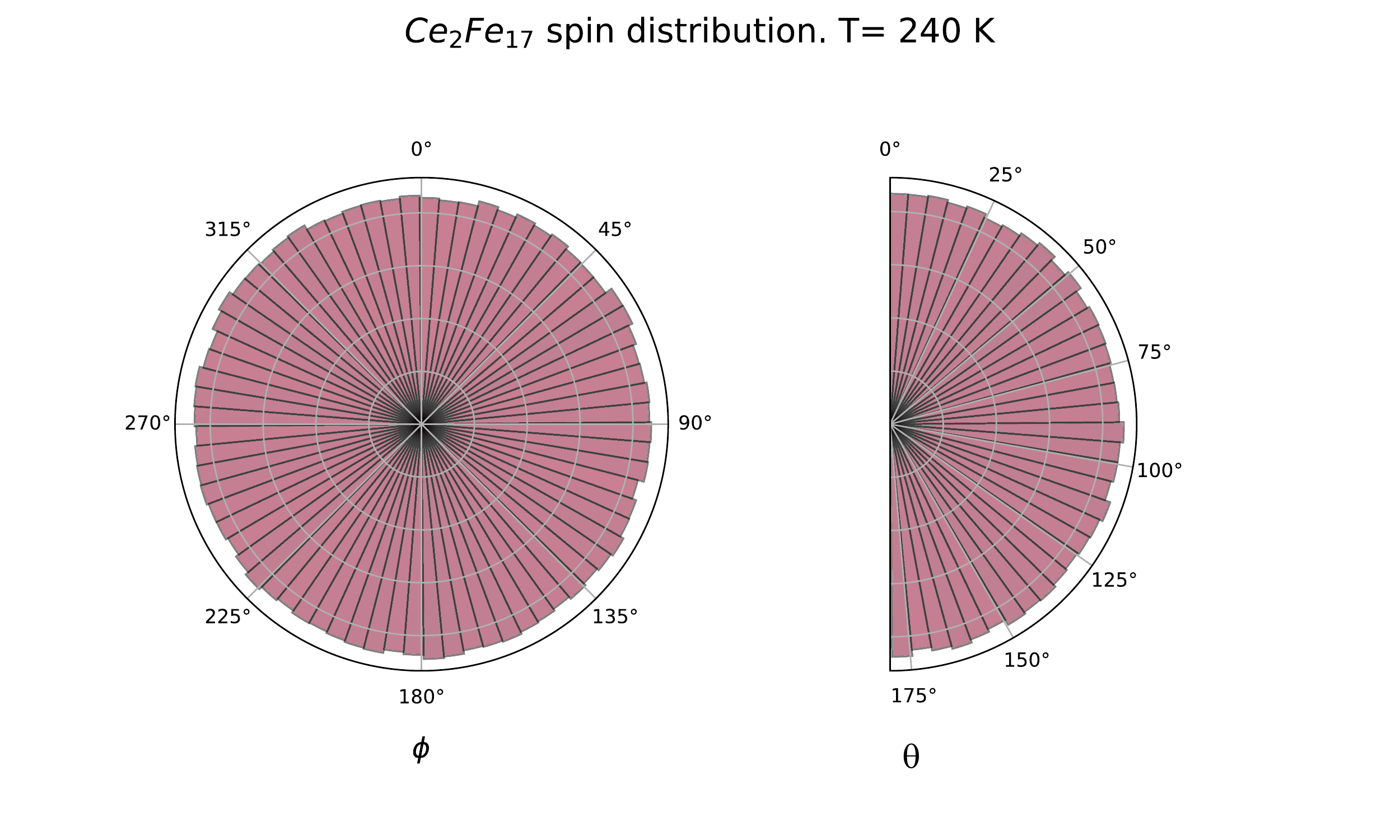}
\end{subfigure}

\caption{(Color online) Distribution of spin directions (in spherical coordinates) in the non-collinear spin structures of Ce$_2$Fe$_{17}$ calculated with UppASD (V = 258.17  $\mathrm{\mathring{A}}^3$ \cite{TEPLYKH2004E99}). The top two figures  belong to the temperature range below T$_{\rm t1}$ where the spin-spiral state has a significant ferromagnetic component. The bottom left diagram describes the spin-spiral state in the region between  T$_{\rm t1}$ and T$_{\rm t2}$ which has a zero total magnetic moment. The figure at the bottom right belongs to the paramagnetic state above T$_{\rm t2}$.}
 \label{fig:angle}
\end{figure*}

In Fig.~\ref{fig:angle} we show the angle-distribution of spins for several temperatures. The top two figures corresponding to the temperatures below $T_{\rm t1}$ clearly demonstrate the significant z-component of magnetization with a preferable direction of the total magnetic moment. The figure at the bottom left shows the spin distribution in the temperature region between $T_{\rm t1}$ and $T_{\rm t2}$ where the net magnet moment is equal to zero. Lastly, the diagram at the bottom right of Fig.~\ref{fig:angle} corresponds to the paramagnetic region. When looking at Fig.~\ref{Cv}, we can see that our theoretical results are in good agreement with the experimental data outlined in Fig.~\ref{Exp}. 

The low-temperature spin structure (Fig.~\ref{SpinUpp}) was found to contain a combination of several spin spirals. The wave vectors present in the spin structure were identified from the static structure factor $S(\mathbf{q})$. In addition to the ferromagnetic component, significant spectral weight was also found for $q$-vectors parallel to {\it c} and equal to 0.048, 0.01, and 0.058 $\mathrm{\mathring{A}}^{-1}$. The first of the values is in a good agreement with the experimental value of 0.0435 $\mathrm{\mathring{A}^{-1}}$ observed by Fukuda {\it et al} \cite{Fukuda1999} for the temperatures below 125 K. Both Fukuda {\it et al} \cite{Fukuda1999} and Givord {\it et al} \cite{Givord} report the value of 0.037 $\mathrm{\mathring{A}^{-1}}$ for the wave vector in the temperature region 125 K $< T <$ 210 K.

The magnetic structure between the two transition temperatures was found to be more difficult to analyze. The spin-spirals present below T$_{\rm t1}$, including the ferromagnetic component, can not be identified above T$_{\rm t2}$. While there is a finite spectral weight for certain other wave-vectors $\mathbf{q}$ in the temperature range, a rather large amount of numerical noise keeps us from further conclusions regarding the magnetic structure between T$_{\rm t1}$ and T$_{\rm t2}$ . 

It is worth mentioning, that for the lowest reported experimental volume (V = 255.77 $\mathrm{\mathring{A}}^3$ \cite{767}) ASD simulations based on GGA input (4{\it f} in the core) result in a low-temperature spin-spiral state similar to the simulations based on LDA data for higher volumes.

\section{Discussion and conclusions}

Despite the large number of reported experimental data for Ce$_2$Fe$_{17}$ compound, the amount of existing theoretical works is rather scarce.   All the previous  computational studies report the FM ground state of this material while considerably overestimating its magnetic moment. Experimental investigations, on the other hand, have not come to a consensus - both FM and AFM states are reported for the low-temperature region. 

We started our investigation by performing  DFT and DFT {\it + U} calculations to determine the ground state of Ce$_2$Fe$_{17}$. One of the difficulties in the current case is treating the 4{\it f} electron of Ce which depends on the degree of localization of  4{\it f} states \cite{Coey}. Deviation from well-localized trivalent core state configuration is quite common for the compounds of Ce. Unlike the other elements in lanthinide series, 4{\it f} orbitals of Ce are sufficiently extended to be capable of significant overlap with the conduction-electron orbitals of the neighboring atoms. The choice of the valence-core configuration is, hence, crucial for the electronic structure and the material properties \cite{Hybrid}.

Experimental investigations report the mixed-valence state of Ce in Ce$_2$Fe$_{17}$, the magnetic moment is determined to be of 5{\it d} character \cite{Miraglia}. The hybridization function calculations we performed likewise shows the intermediate state between the materials with itinerant 4{\it f} electrons and compounds with the localized 4{\it f} shell. Hence, we performed the DFT calculations for two cases, treating the {\it f}-electron as a core and a valence state. Both  calculations (in GGA approximation) overestimate the total magnetic moment by 30-40 \%. Treating 4{\it f} as a valence state, however, gives the unit cell volume which closely reproduces experimental data; GGA {\it + f in the core} calculation overestimates the most common experimental volume of 258 $\mathrm{\mathring{A}}^3$ by about 4\%. 
LDA fails in the structure relaxation of Ce$_2$Fe$_{17}$ with a relaxed u.c. volume underestimating the experimental one by 15\% and the total magnetic moment of the relaxed unit cell 80\% below experimental. However, if applied to the experimental unit cell (or the structure relaxed in GGA), LDA produces the total magnetic moment in good agreement with the experimental one.
It has been reported before \cite{PhysRevB.79.085104} that using PBE functional produces better results for structure relaxation of many materials containing 3{\it d} elements, while LDA sometimes gives a better result for the magnetic state \cite{PhysRevB.94.024423}. Ce$_2$Fe$_{17}$ is another case where the ground state needs to be treated with the LDA. This fact is most likely due to the GGA not catching the intermediate nature of 4{\it f} state, as shown in Appendix~\ref{sec:hybgga}.



ASD simulations performed based on the GGA magnetic configuration, as might be expected, predict the FM state of Ce$_2$Fe$_{17}$ up to the T$_{\rm C}$ of about 600 K that considerably overestimates the experimental value. At the same time,  LDA input results in a non-collinear spin structure with a significant ferromagnetic component which becomes non-magnetic at T$_{\rm C}$ = 93 K, quite close to the experimental value. ASD simulations closely reproduce the experimental direction and size of the wave vectors of the non-collinear spin structure.
Therefore, we believe that the low-temperature state of Ce$_2$Fe$_{17}$ (named FM in many experimental works) is non-collinear with a strong FM component, i.e. that it is a helical phase. 


\section{Acknowledgement}

The authors would like to acknowledge the support of the Swedish Foundation for Strategic Research, the Swedish Energy Agency, the Swedish Research Council, The Knut and Alice Wallenberg Foundation, eSSENCE, STandUPP the CSC IT Centre for Science, and the ERC (synergy grant FASTCORR, project 854843). 
In addition, we acknowledge support from the Swedish National Infrastructure for Computing (SNIC) for the computation resources (projects SNIC 2020/8-34 and SNIC 2020/1-20). O. Yu. V. acknowledges the support of Sweden's Innovation Agency (Vinnova). 



\section{Competing interests}

The authors declare no competing interests.

 \bibliographystyle{elsarticle-num} 
\bibliography{cas-refs}

\onecolumn
\appendix

\newpage

\section{Overview of the available experimental and theoretical data for Ce$_2$Fe$_{17}$.}\label{sec:details}

In this section, we present the existing experimental and theoretical data available for Ce$_2$Fe$_{17}$; the main details of the magnetic state and structure parameters are outlined in Table \ref{table:1} and presented in the diagram in Fig.~\ref{Exp}.

Two crystalline forms of Ce$_2$Fe$_{17}$ were reported by Buschow and Wieringen \cite{Buschow}: the hexagonal, Th$_2$Ni$_{17}$ ($\alpha$-Ce$_2$Fe$_{17}$), and the rhombohedral, Th$_2$Zn$_{17}$ ($\beta$-Ce$_2$Fe$_{17}$) phases. The hexagonal structure could not be obtained as single-phase \cite{Buschow}; the rhombohedral structure is the one considered in all the following experimental research. 

Buschow and Wieringen \cite{Buschow} observed the  ferromagnetic (FM) (with a moment of 1.80 $\mu_B$ per Fe atom) to antiferromagnetic (AFM) transition near T=70 K and an antiferromagnetic to paramagnetic (PM) transition close to T$_{\text{N}}$=270 K, for the compound the $\beta$-Ce$_2$Fe$_{17}$. From the lattice constants, Ce is in this work suggested to be tetravalent.

Two samples of Ce$_2$Fe$_{17}$ were prepared in Ref.~\cite{kpfu} by two different methods demonstrating different magnetic behavior. One of the samples has a FM to AFM transition at T=22 K, with another transition between different helicoidal states at 75 K and 109 K. A higher temperature of the FM to AFM transition (T=94 K) is observed in the second sample.  The differences in sample preparation are the crucibles (a semi-gravitational copper crucible and a crucible made of Al$_2$O$_3$) and the annealing times (8 and 300 h, both at 1000$^{\circ}$C). The sample with the smaller temperature for the FM to AFM transition, has smaller lattice parameters. The authors of Ref.~\cite{kpfu} report M{\"o}ssbauer spectra that show the regions with AFM ordering in FM samples. Furthermore, a temperature for the AFM to FM transition of T=94 K was reported, together with a N{\'e}el temperature, T$_{\text{N}}$=206 K, by Kuchin {\it et al} \cite{KUCHIN20007}.

\begin{table*}[h]
\caption{Overview of the available experimental and theoretical data on Ce$_2$Fe$_{17}$ (Reported ground state, Curie, Ne\'el, and transition temperatures, magnetization per Fe atom (or per f.u.), lattice parameters, lattice data measurement temperature). Several lines per paper correspond to different samples.  `Sat' stands for saturation moment, `calc' marks theoretical calculations, 'IF' notes induced FM state. * gives the real composition of the samples in \cite{TEPLYKH2004E99}. }
\begin{tabular}{c c c c c c c c c c} 
 \hline \hline
 Reference & Ground & T$_{\text{C}}$ & T$_{\text{N}}$ & T$_{\text{t}}$ &  M (per Fe, & a & c & V & T\\ 
  & state & K & K & K & or f.u., $\mu_B$ & $\mathrm{\mathring{A}}$ & $\mathrm{\mathring{A}}$ & $\mathrm{\mathring{A}}^3$ & \\
 \hline
Buschov {\it et al} \cite{Buschow} & FM & 70 & 270 &  & 1.8 (sat) & 8.49 & 12.416 & 258.3 & room \\ 
Naumov {\it et al} \cite{kpfu} & FM & 22 & 212 & 75,109 &  & 8.49 & 12.408 & 258.2 & room \\ 
 & FM & 94 & 207.8 &  &  & 8.495 & 12.415 & 258.6 & room \\
Kuchin {\it et al} \cite{KUCHIN20007} & FM & 94 & 206 & &  & 8.492 & 12.408 & 258.3 & room \\
Teplych {\it et al} \cite{TEPLYKH2004E99} & & & & & & & & & \\
Ce$_2$Fe$_{17.31}$* & FM & 20 & 203 & 102 & 29.6/f.u. (4 K)  & 8.4897 & 12.4083 & 258.17 & room \\
Ce$_2$Fe$_{17.23}$* & FM & 47 & 210 & 101 &  & 8.4923 & 12.4098 & 258.36 & room \\
Ce$_2$Fe$_{17.19}$* & FM & 85 & 203 & & & 8.4949 & 12.4112 & 258.52 & room \\
Ce$_2$Fe$_{16.89}$* & FM & 116 & 205 & & 29.8/f.u. (4 K) & 8.5027 & 12.4180 & 259.16 & room \\
Givord {\it et al} \cite{Givord} & fan &  & 225 & 90 & 1.55 (140 K) & & & & \\
Janssen {\it et al} \cite{Janssen} & helical & & 215 & 125 & 1.7 (5 K IF) & & & & \\
Kreyssig {\it et al} \cite{Supercell} & AFM & & 118 & &  & 8.489 & 12.41 & 258.2 & room \\
Janssen {\it et al} \cite{Janssen2007} & AFM & & 208 & 124 &  & 8.489 & 12.41 & 258.2 & room \\
Makihara {\it et al} \cite{MAKIHARA2003663} & fan & & & &  & 8.489 & 12.468 & 259.4 & 8 K \\
 & & & & &  & 8.494 & 12.425 & 258.8 & room \\
 & helical & & & &  & 8.467 & 12.44 & 257.4 & 8 K \\
 & & & & &  & 8.487 & 12.41 & 258.0 & room \\
Isnard {\it et al} \cite{Miraglia} & helical & & 225 &  & 30.46/f.u. (4 K)  & 8.49 & 12.413 & 258.3 & room \\
Koyama {\it et al} \cite {KOYAMA2001168} & helical & & 210 & 125 & 30.5/f.u. (sat)  &  &  &  &  \\
Hautot {\it et al} \cite {Hautot} & fan & & 225 & 90 & 1.49 (4 K) &  &  &  &  \\
Pandey {\it et al} \cite {PhysRevApplied.13.034039} (calc) & FM & & & & 37.43/f.u. &  &  &  &  \\
S\"ozen {\it et al} \cite {PhysRevMaterials.3.084407} (calc) & FM & & & & 37.46/f.u. & 8.403 & 12.641 & 257.7 &  \\
Lukoyanov {\it et al} \cite {Lukoyanov2007} (calc) & FM & & & & 35.98/f.u. & & & &  \\
 \hline \hline
\end{tabular}
\label{table:1}
\end{table*}

Seven samples were considered in Ref.\cite{TEPLYKH2004E99} to determine the correlation between their crystal structures and magnetic states. The FM to AFM transition temperature of the samples changes drastically from 20 K to 116 K; while T$_{\text{N}}$ and magnetic moment remain mostly unchanged.


Some experiments, however, show the absence of the FM phase. Givord and Lemaire \cite{Givord} reported a fan magnetic order of Ce$_2$Fe$_{17}$ below 90 K and a helical magnetic order between T$_{\text{C}}$=90 K and T$_{\text{N}}$=225 K. The fan magnetic structure is parallel to the basal plane with the fan angle of 40$^{\circ}$ at 4.2 K. The helical spin configuration at 140 K is characterized by the rotational axis parallel to {\it c}-axis and magnetization in the basal plane rotating by an angle of 27$^{\circ}$ from plane to plane, and a wave vector $q=0.037$ $\mathrm{\mathring{A}}$; the mean iron moment is 1.55 $\mu_B$.

In Ref.\cite{MAKIHARA2003663} the temperature and magnetic field dependence of the lattice parameter were investigated for two samples of  Ce$_2$Fe$_{17}$ with two different ground states - the fan, and the helical spin structures. They note that by applying a magnetic field of 5 T the helical spin state can be transformed into the fan spin structure. Anisotropy of the exchange interactions was determined from the thermal change in {\it a} and {\it c} lattice parameters. Magnetoelasticity of Ce$_2$Fe$_{17}$ was also studied in \cite{ANDREEV200043}. Similar to other reports, the authors of this work, note that the magnetic properties of the compound are sample-dependent, especially for the low-temperature range.

Janssen {\it et al} \cite{Janssen} report the giant magnetoresistance in the rhombohedral Th$_2$Zn$_{17}$-type structure of Ce$_2$Fe$_{17}$. They also observe two transition temperatures, 125 K, and 215 K. These are considered to be first the critical temperature at which the magnetic structure changes from one AFM order to another and secondly the Ne{\'e}l  temperature. The authors also state that the first AFM order is not a fan structure  because this phase does not have any FM component. Induced FM order is reported in the higher fields in which the average Fe moment is 1.70 $\mu_B$ at 5 K. 

A helical spin structure with a wave vector $q=0.0372$ $\mathrm{\mathring{A}}$ parallel to the $c$-axis is reported by Fukuda {\it et al} \cite{Fukuda1999} in the temperature region between T$_{\bf C}$=125 K and T$_{\bf N}$=210 K. A modified helix is observed below T$_{\bf C}$ with $q=0.0435$ $\mathrm{\mathring{A}}$ and a super-structured magnetic cell with $c$ parameter twice as large as in the chemical unit cell.

In Ref. \cite{Janssen2007} authors report Ce$_2$Fe$_{17}$ to be AFM with two transition temperatures, T$_{\text{N}}$=208 K and T$_{\text{t}}$=124 K, where T$_{\text{t}}$ is a second magnetic-order transition to another AF phase. They also present the magnetic phase {\it H-T} diagram of Ce$_2$Fe$_{17}$.

Kreyssig {\it et al} \cite{Supercell} report a superstructure that was observed in Ce$_2$Fe$_{17}$  below 118 K as the AF ordered state, which can be constructed by the doubling of a unit cell in the {\it c}-direction with the same space group as the room-temperature structure. 

A significant decrease in T$_{\text{C}}$ has been observed when the hydrostatic pressure was applied in \cite{pressure1,pressure2}. Prokhnenko {\it et al} report \cite{pressure2} that the FM state can be suppressed by pressures higher than 3 kbar and that it is replaced by a new AFM phase. Koyama {\it et al} \cite {KOYAMA2001168} have performed magnetization measurements on Ce$_2$Fe$_{17}$ under high pressures and magnetic fields. There are two transition temperatures T$_{\text{N}}$=210 K and T$_{\text{t}}$=125K observed at the ambient pressure of 0 GPa with the helical structure below T$_{\text{t}}$. 

Saturation  magnetization  decreases  from 30.5 $\mu_B$/f.u. at 0 GPa to 29.5 $\mu_B$/f.u. at 0.8 GPa. The authors report at least seven magnetic phases found in the {\it B-T} diagram.
It was also concluded that hybridization between the Fe 3{\it d}- and Ce 4{\it f}-electron states is extremely important for the magnetic state of Ce$_2$Fe$_{17}$.

Coey {\it et al} \cite{Coey} discuss the hybridization between Ce 4{\it f} and Fe 3{\it d} electron states, and it was suggested that in this material there is no pure, localized 4{\it f}$^1$ configuration of Ce. This analysis is in line with previous, more general discussions on the nature of the 4f states of Ce-Fe compounds \cite{PhysRevLett.60.2523}.
The decrease in Ce$_2$Fe$_{17}$ u.c. volume within a series of isostructural rare-earth intermetallics is attributed to 4{\it f}  electrons behaving as itinerant electrons with a heavy mass. Fujii {\it et al} \cite{Fujii} showed that the nitrogenation of Ce$_2$Fe$_{17}$ causes the conversion of itinerant 4{\it f}-electron state into a  more localized but valence fluctuating 4{\it f}-electron state. A similar effect seems to be observed upon hydrogenation \cite{ISNARD1992157}. This change in the 4{\it f} electron character as the effect of interstitial hydrogen and nitrogen is also described in Ref.\cite{FUJII199510}. 

Isnard {\it et al} \cite{Miraglia} investigate the change in the values of Ce valence in Ce$_2$Fe$_{17}$H$_x$ with the addition of hydrogen based on the absorption and magnetic circular x-ray dichroism experiments. They report a mixed valence state ($v$=3.3) of Ce in Ce$_2$Fe$_{17}$. The magnetic moment carried by the Ce atoms is determined to be of 5{\it d} character.  The 4{\it f} state in all Ce$_2$Fe$_{17}$H$_x$ compounds is reported to be delocalized. As x changes from 0 to 5  a noticeable increase in the u.c. volume from 258.3 to 271.9 $\mathrm{\mathring{A}}^3$ is observed along with the increase in magnetization from 30.46 to 35.35 $\mu_B$/f.u. along with a suggested change in Ce valence from 3.33 to 3.26.

Valeanu  {\it et al} state in Ref.\cite{Valeanu} that iron magnetism shows a gradual transition from an itinerant behavior in R$_2$Fe$_{17}$ to a rather localized one in R$_2$Fe$_{17}$C$_x$, where R is a RE element.
A similar effect is reported in Ref.\cite{Isnard_1996} for Ce$_2$Fe$_{17}$N$_x$. With an increase in x from 0 to 3, the volume increases to 280.8 $\mathrm{\mathring{A}}^3$ as the magnetization changes from 29.7 to 37.7 $\mu_B$/f.u. and the transition temperature increases from T$_{\bf N}$=225 K to T$_{\bf C}$=728 K. According to the authors,  Ce$_2$Fe$_{17}$ demonstrates a helimagnetic structure below the N{\'e}el temperature of  225  K  while the nitrogenated sample of  Ce$_2$Fe$_{17}$N$_3$ exhibits  a  ferromagnetic-like  behaviour  below  T$_{\bf C}$=728 K.

The effect of Ga and H substitution on the degree of localization of the 4{\it f} states of cerium and the magnetic state of Ce$_2$Fe$_{17-x}$Ga$_x$H$_y$ was also considered in \cite{ISNARD1997198} with X-ray absorption spectroscopy combined with magnetization measurements. The authors report that substitution of iron by gallium as well as the addition of hydrogen lead to a localization of the 4{\it f} states of Ce in Ce$_2$Fe$_{17-x}$Ga$_x$ with the values of $v$=3.26 for Ce$_2$Fe$_{16}$Ga, $v$=3.22 for Ce$_2$Fe$_{15}$Ga$_2$, $v$=3.16 for Ce$_2$Fe$_{16}$GaH$_5$. Ce 5{\it d} magnetic moment in Ce$_2$Fe$_{15}$Ga$_2$ is shown to couple antiparallel to the 3{\it d} magnetic moment of Fe and is estimated to be 0.35 $\mu_B$.

M{\"o}ssbauer spectral study of the magnetic properties of Ce$_2$Fe$_{17}$H$_x$ with x=0,1,2,3,4, and 5 was performed by Hautot {\it et al} \cite{Hautot} taking into account the fan and helical magnetic structure of Ce$_2$Fe$_{17}$. The authors found that  iron magnetic moments in Ce$_2$Fe$_{17}$ range from 0.91 $\mu_B$ to 2.13 $\mu_B$ depending on the Fe site with the average iron magnetic moment of 1.49 $\mu_B$.

A magnetic study of Ce$_2$Fe$_{17}$H$_x$ is also presented in Ref. \cite{Miraglia}. The changes in transition temperature from T$_{\bf N}$ = 225 to T$_{\bf C}$ = 444 K and magnetic moment from 30.46 to 35.35 $\mu_B$/f.u. are noted as {\it x} goes from 0 to 5. At the same time Ce valence was suggested to shift from 3.33 to 3.26.

In Ref.\cite{Lukoyanov2007} the authors combined the experimental and theoretical studies of Ce$_2$Fe$_{17}$. Using a Tight-Binding Linear Muffin-Tin Orbitals Atomic-Sphere Approximation (TB-LMTO-ASA) program package \cite{Andersen}, without taking into account spin-orbit coupling, they obtained a total moment of 35.98 $\mu_B$/f.u. (-0.61 $\mu_B$ at Ce 6{\it c} site, 2.47 $\mu_B$ at Fe 6{\it c} site, 2.06 $\mu_B$ at Fe 9{\it d} site, 2.25 $\mu_B$ at Fe 18{\it f} site, and 2.15 $\mu_B$ at Fe 18{\it h} site).

Moreover, in their recent theoretical work, Pandey and Parker \cite{PhysRevApplied.13.034039} performed first-principles DFT + {\it U} ($U_{eff}=3$ eV on the Ce 4$f$ orbital) calculations using the general potential linearized augmented plane-wave (LAPW) method plus local orbitals implemented in WIEN2K code \cite{WIEN2K} and generalized gradient approximation (GGA) of Perdew, Burke, and Erzerhof \cite{PBE} to investigate Zr doping of Ce$_2$Fe$_{17}$ and Ce$_2$Co$_{17}$. They model Ce$_2$Fe$_{17}$ as a ferromagnet and obtain a total magnetic moment of 38.04 $\mu_B$/f.u. (-0.66 $\mu_B$ at Ce 6{\it c} site, 2.56 $\mu_B$ at Fe 6{\it c} site, 2.05 $\mu_B$ at Fe 9{\it d} site, 2.36 $\mu_B$ at Fe 18{\it f} site, and 2.22 $\mu_B$ at Fe 18{\it h} site) for the experimental lattice parameters.

In Ref.\cite{PhysRevMaterials.3.084407} the authors investigated the properties of Ce-based magnetic materials. DFT calculations without Hubbard {\it U} parameter were performed for Ce-Fe binary alloys with 5{\it s}$^2$5{\it p}$^6$4{\it f}$^1$5{\it d}$^1$6{\it s}$^2$ valence configuration of Ce in GGA by Perdew and coworkers \cite{PBE}. Their collinear spin-polarized calculation with Vienna {\it ab initio} simulation package VASP \cite{vasp1,vasp2,vasp3,vasp4} using the projector-augmented wave (PAW) \cite{PAW} method gives the following magnetic moments for Ce$_2$Fe$_{17}$: 2.59 $\mu_B$ at Fe 6{\it c} site, 2.06 $\mu_B$ at Fe 9{\it d} site, 2.39 $\mu_B$ at Fe 18{\it f} site, 2.23 $\mu_B$ at Fe 18{\it h} site, and -0.81 $\mu_B$ at Ce 6{\it c} site.

\section{Hybridization function in LDA and GGA} \label{sec:hybgga}

\begin{figure}[h!]
\begin{subfigure}[b]{0.47\textwidth}
\includegraphics[scale=0.31]{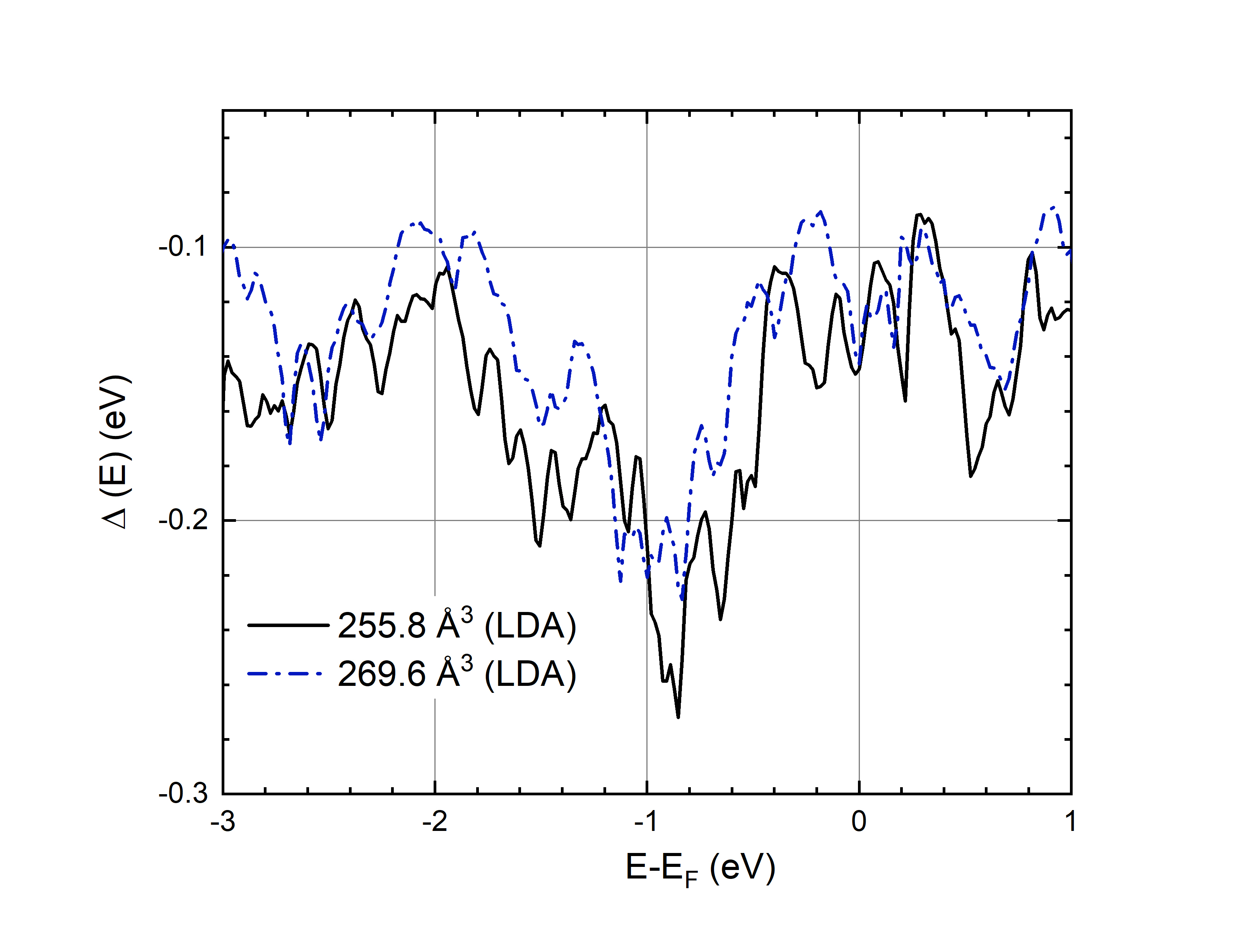} 
\label{Hyb}
\end{subfigure}
\begin{subfigure}[b]{0.47\textwidth}
\includegraphics[scale=0.31]{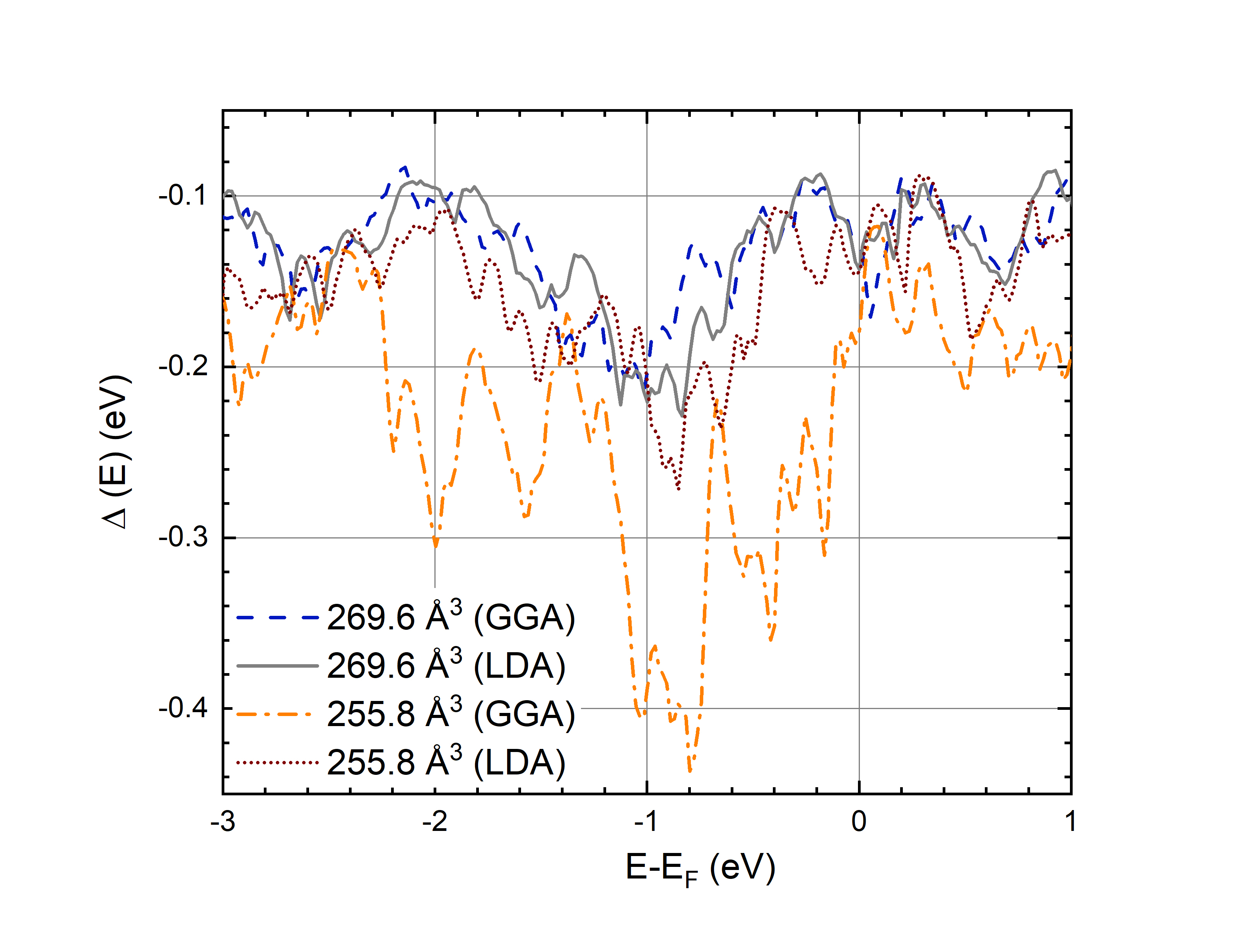}
\label{Hyball}
\end{subfigure}
 \caption{(Color online) The hybridization function $\Delta (E)$ of Ce$_2$Fe$_{17}$, as a function of energy relative to the Fermi energy for the smallest and the largest experimental unit cells available in ICSD, calculated in the framework of LDA (left) and comparison of the hybridization functions calculated within LDA and GGA (right)}
\label{Hybridization}
\end{figure}

Fig.~\ref{Hybridization} (left) shows the change of hybridization function calculated within the framework of LDA with the volume.

In addition to the hybridization function, $\Delta (E)$, calculated in LDA, we performed the similar calculations in GGA. The hybridization function was calculated for the largest and the smallest experimental unit cells found in ICSD, see the right panel of Fig.~\ref{Hybridization}. 

We can see, that unlike the case of LDA in GGA the absolute value of the peak situated about 1 eV below E$_F$ changes significantly from 0.44 eV for 255.8 $\mathrm{\mathring{A}}^3$ to 0.22 eV for 269.6 $\mathrm{\mathring{A}}^3$. The u.c. volume observed in most of the experiments is 258.2 $\mathrm{\mathring{A}}^3$. Hence, for most of the experimental unit cells GGA doesn't catch the intermediate nature of the 4f state (see discussion in Sec.~\ref{sec:main}). This fact might explain the failure of GGA to reproduce the magnetic state of  Ce$_2$Fe$_{17}$ correctly. 

\section{LDA + U calculations and DOS} \label{sec:DOS}

\begin{figure}[h!]
\begin{subfigure}[b]{0.47\textwidth}
\includegraphics[scale=0.37]{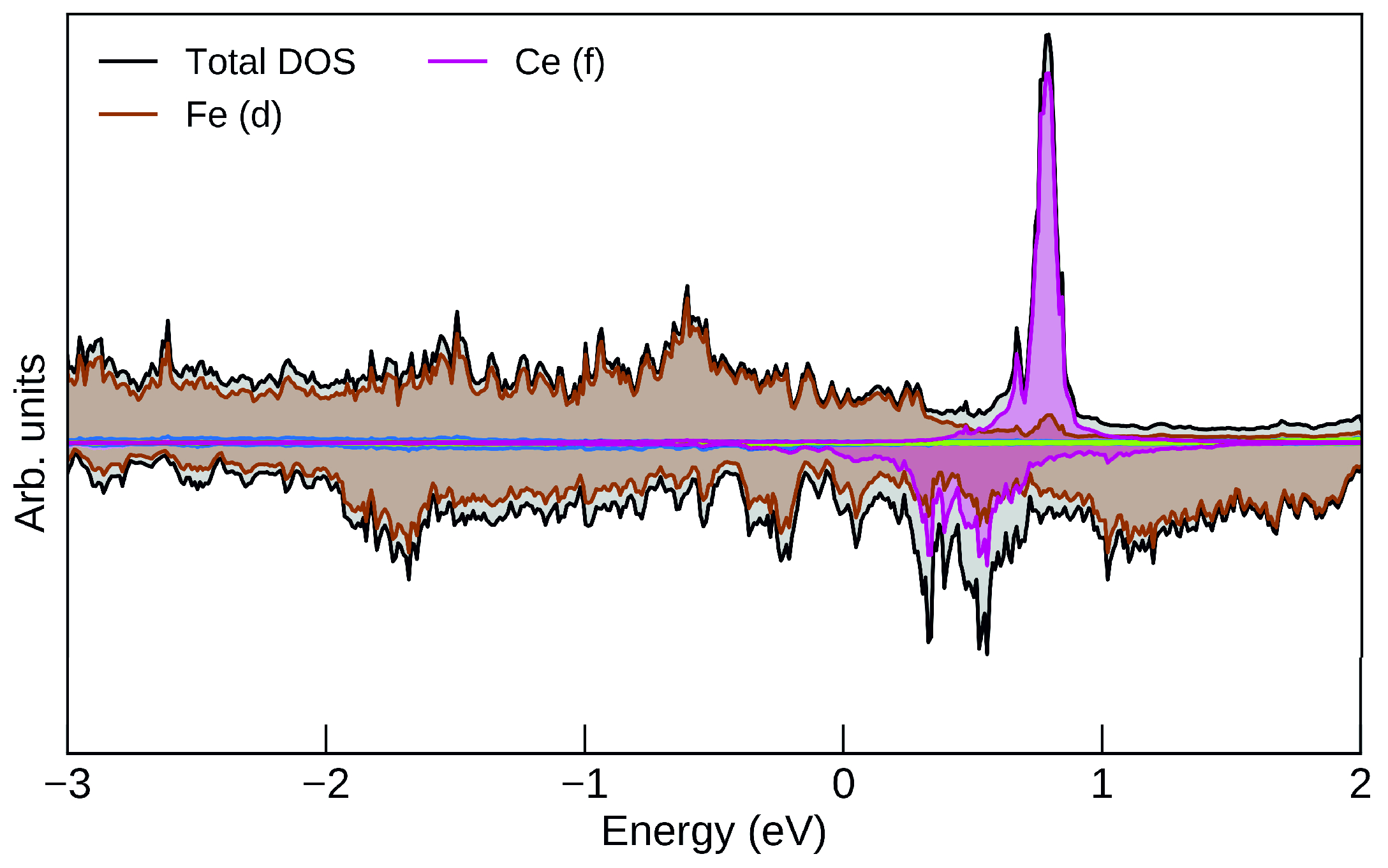} 
\label{dos_LDA}
\end{subfigure}
\begin{subfigure}[b]{0.47\textwidth}
\includegraphics[scale=0.37]{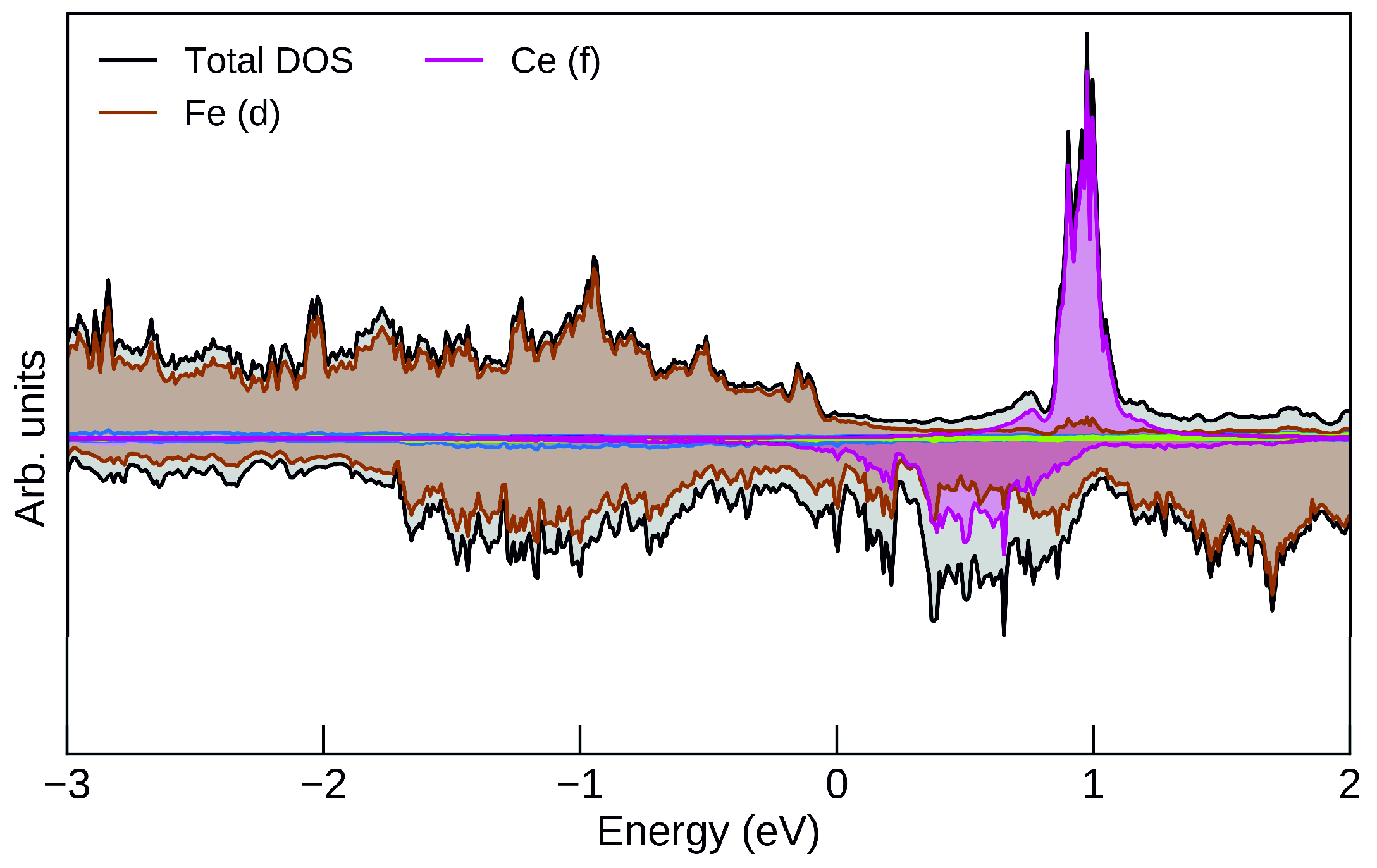}
\label{dos_GGA}
\end{subfigure}
\begin{subfigure}[b]{0.47\textwidth}
\includegraphics[scale=0.27]{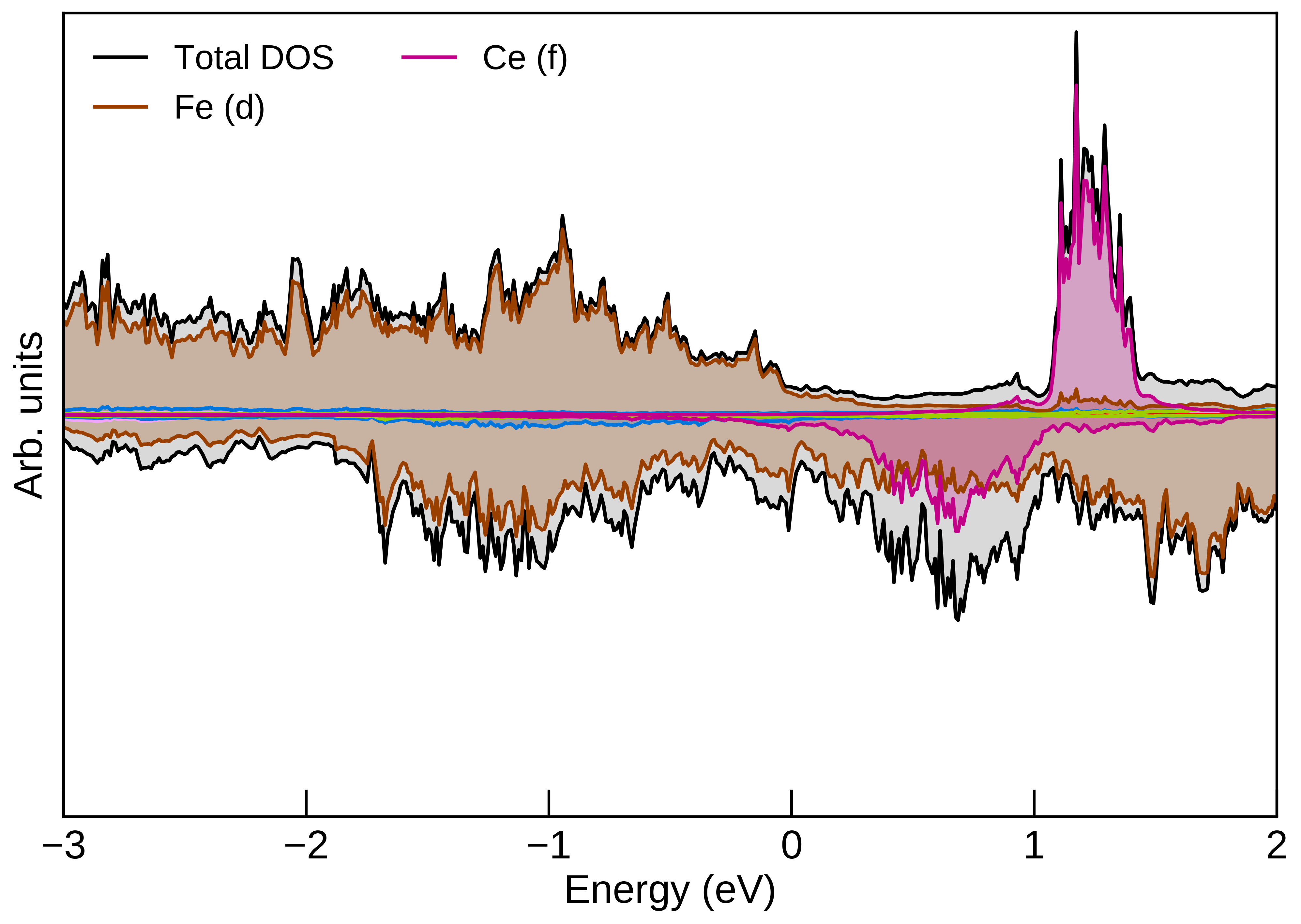}
\label{dos_U_3}
\end{subfigure}
 \caption{(Color online) Spin polarized DOS of Ce$_2$Fe$_{17}$ calculated for a typical experimental unit cell in LDA (top), GGA (middle), and DOS calculated with 4f states treated with the GGA+U approximation (bottom). In all plots, spin-up states are shown as positive while spin-down as negative.}
\label{DOS}
\end{figure}

We used several values of $U-J$ for the 4{\it f} electron of Ce; as is shown in Table \ref{table:2} this approach makes little change to the total magnetic moment and crystal structure parameters of Ce$_2$Fe$_{17}$. The reason for this insensitivity can be understood by investigating the DOS of Ce$_2$Fe$_{17}$, Figure \ref{DOS}. All of the spin-up 4{\it f} states of Ce and most of the spin-down states are empty with the spin-up peak lying $\sim$ 1 eV above the Fermi level. The DOS is calculated both in GGA and LDA. For the former, the spin-up channel of Fe is completely filled and the spin-down channel is partially occupied. The main difference between the two lies in the spin-up {\it d}-states of iron which are not fully filled in the case of LDA. Different values of {\it U} produce only a slight shift in the 4{\it f} spin-up peak.

\section{Fixed spin moment calculations} \label{sec:fixed}

\begin{figure}[h!]
 \centering
 \includegraphics[scale=0.3]{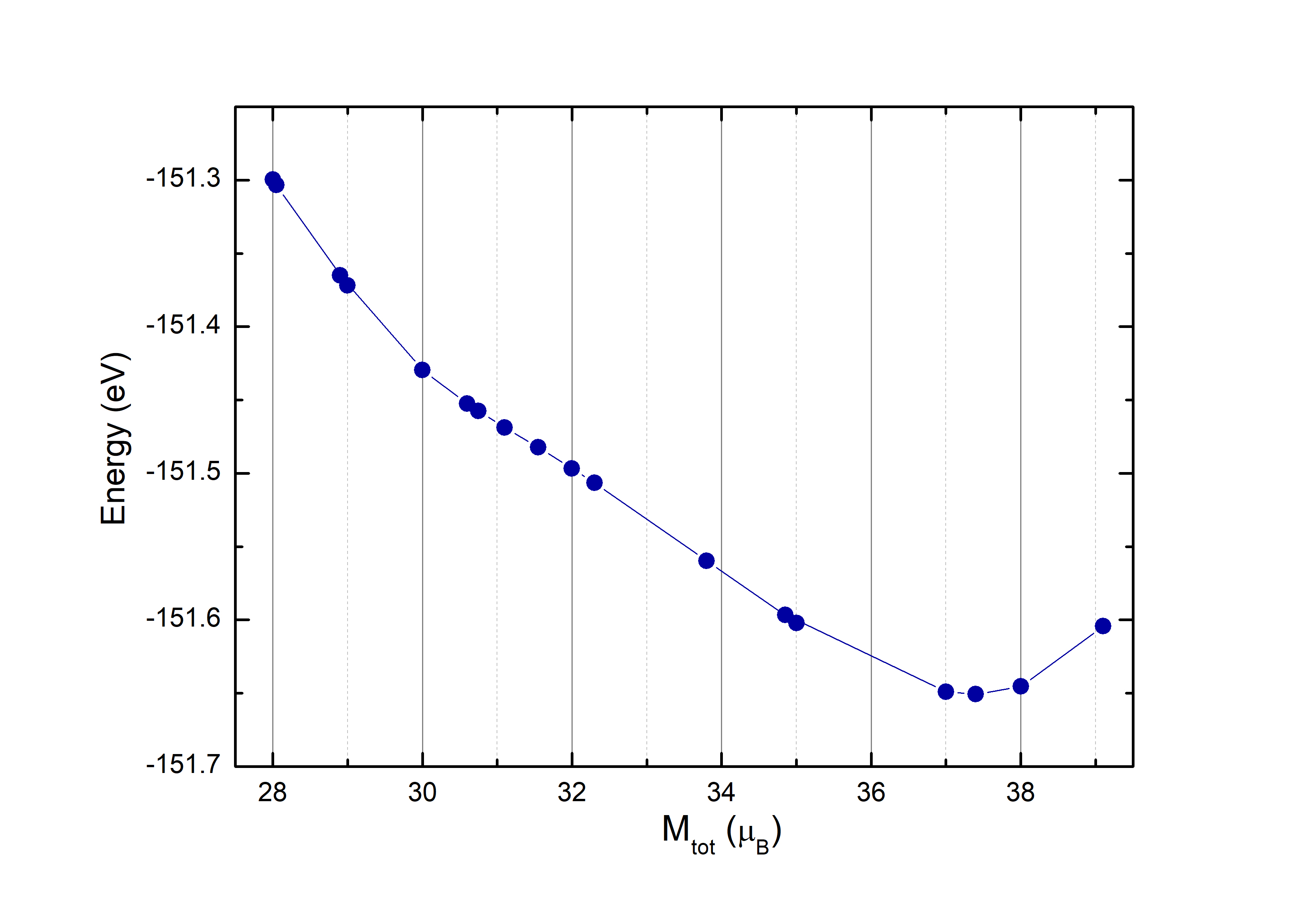}
 \caption{(Color online) Total energy calculated for different values of the fixed spin magnetic moment. The 4f states are treated as itinerant, and the GGA functional was employed.}
 \label{FixedM}
\end{figure}

Fixed spin moment (FSM) calculations are sometimes utilized to investigate the existence of the local and global energy minima, as function of the total magnetic moment \cite{FixM}. The method was first introduced by Schwarz and Mohn \cite{Schwarz_1984}. In the case of LaFe$_{13-x}$Si$_x$ \cite{PhysRevB.76.092401}, for instance, FSM curves show a shift of the system's magnetic state from one local minimum to another with the change of the lattice parameter. We performed similar calculations for Ce$_2$Fe$_{17}$, by changing the total moment within a region of values containing the experimental magnetic moment of about 29 $\mu_B$/f.u. and a calculated magnetic moment of 38 $\mu_B$/f.u. In these calculations the 4{\it f} electrons were treated as valence (GGA, without a {\it U}-term). The results are shown in Fig. \ref{FixedM}. No pronounced minimum is observed around the experimental value, however, the curve does demonstrate a slight deviation from the monotonic decrease in the region around 30 $\mu_B$/f.u. The global minimum is located at the value of 37.5 $\mu_B$/f.u. 

\section{Details of ASD simulations in GGA} \label{sec:spinGGA}

\begin{figure}[h!]
 \centering
 \includegraphics[scale=0.25]{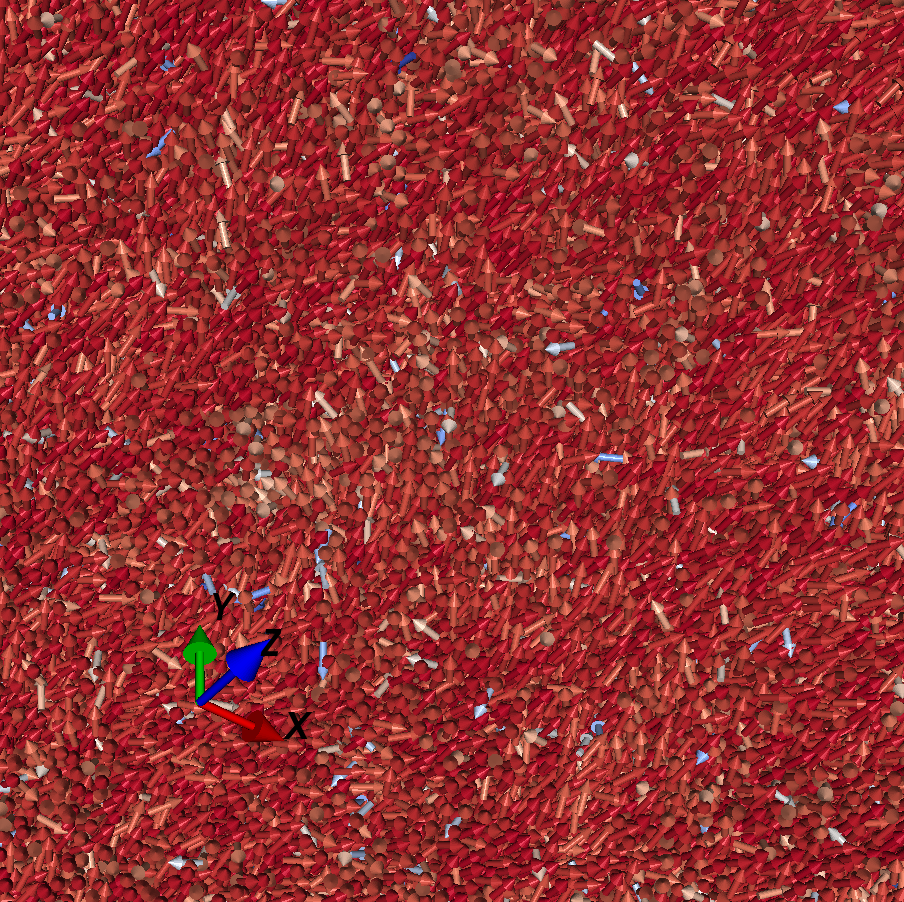}
 \caption{(Color online) Magnetic configuration of Ce$_2$Fe$_{17}$ calculated with ASD simulations based on exchange parameters calculated in GGA (T = 100 K, for the experimental structure with volume V = 255.77 $\mathrm{\mathring{A}}^3$ \cite{767}).}
 \label{FMsnap}
\end{figure}

ASD simulations, based on the GGA exchange parameters and magnetic moments, result in FM spin order up to the high temperatures of 400-500 K depending on the u.c. volume. This result contradicts all the experimental findings. Fig.~\ref{FMsnap} shows the simulated spin structure for the experimental u.c. with V = 255.77 $\mathrm{\mathring{A}}^3$ \cite{767} at T = 100 K.

\end{document}